\documentclass[12pt]{article}   
\usepackage{geometry,bbm}   
\usepackage[toc,page]{appendix}             		
\usepackage{graphicx}
\usepackage[usenames,dvipsnames]{color}
\usepackage{slashed}					
\usepackage{amssymb,amsmath}
\usepackage{hyperref}
\usepackage{mathtools}

\newcommand{\nn}{\nonumber}

\newcommand{\bi}{\begin{itemize}}
\newcommand{\ei}{\end{itemize}}

\newcommand{\ben}{\begin{enumerate}}
\newcommand{\een}{\end{enumerate}}

\def\){\right)}
\def\({\left( }
\def\]{\right] }
\def\[{\left[ }

\newcommand{\be}{\begin{equation}}
\newcommand{\ee}{\end{equation}}

\def\bea{\begin{eqnarray}}
\def\eea{\end{eqnarray}}

\def\bal#1\eal{\begin{align}#1\end{align}}

\def\bald{\begin{aligned}}
\def\eald{\end{aligned}}

\def\beqx{\begin{displaymath}}
\def\eeqx{\end{displaymath}}

\newcommand{\bmat}{\left(\begin{array}}
\newcommand{\emat}{\end{array}\right)}

\def\e{\epsilon}

\def\cn{{\cal N}}

\def\bo{{\raise-.3ex\hbox{\large$\Box$}}}               
\def\face{{\raise.2ex\hbox{$\displaystyle \bigodot$}\mskip-2.2mu \llap {$\ddot
        \smile$}}}                                   
\def\>{\rangle}                                      
\def\<{\langle}                                      

\def\leftrightarrowfill{$\mathsurround=0pt \mathord\leftarrow \mkern-6mu
        \cleaders\hbox{$\mkern-2mu \mathord- \mkern-2mu$}\hfill
        \mkern-6mu \mathord\rightarrow$}        
\def\dvec#1{\vbox{\ialign{##\crcr
        \leftrightarrowfill\crcr\noalign{\kern-1pt\nointerlineskip}
        $\hfil\displaystyle{#1}\hfil$\crcr}}}           

\def\-{\hphantom{-}}

\begin{document}

\begin{titlepage}

\begin{flushright}
SISSA 58/2017/FISI
\end{flushright}
\bigskip
\def\thefootnote{\fnsymbol{footnote}}

\begin{center}
{\LARGE
{\bf
Orientifolds and duality cascades: \vskip 12pt  confinement before the wall

}}
\end{center}

\bigskip
\begin{center}
{\large
Riccardo Argurio$^{1}$ and Matteo Bertolini$^{2,3}$}

\end{center}

\renewcommand{\thefootnote}{\arabic{footnote}}

\begin{center}
\vspace{0.2cm}
$^1$ {Physique Th\'eorique et Math\'ematique and International Solvay Institutes \\ Universit\'e Libre de Bruxelles, C.P. 231, 1050 Brussels, Belgium\\}
$^2$ {SISSA and INFN - 
Via Bonomea 265; I 34136 Trieste, Italy\\}
$^3$ {ICTP - 
Strada Costiera 11; I 34014 Trieste, Italy\\}

\vskip 5pt
{\texttt{rargurio@ulb.ac.be,bertmat@sissa.it}}

\end{center}

\vskip 5pt
\noindent
\begin{center} {\bf Abstract} \end{center}
\noindent
We consider D-branes at orientifold singularities and discuss two properties of the corresponding low energy four-dimensional effective theories which are not shared, generically,  by other Calabi-Yau singularities. The first property is that duality cascades are finite and, unlike ordinary ones, do not require an infinite number of degrees of freedom to be UV-completed.  The second is that orientifolds tend to stabilize runaway directions. These two properties can have interesting implications and widen in an intriguing way the variety of gauge theories one can describe using D-branes.

\vspace{1.6 cm}
\vfill

\end{titlepage}

\newpage
\tableofcontents

\section{Introduction and summary}

D-branes at Calabi-Yau (CY) singularities represent a powerful tool to engineer interesting and rich gauge theory dynamics in string theory. This has applications in many different contexts, ranging from gauge/gravity duality, including models of supersymmetric confining dynamics and dynamical supersymmetry breaking, to string model building, string compactifications and string cosmology.

For a generic CY singularity, it is not simple to know what the D-brane low energy dynamics is. However, there exists a large class of CYs, the so-called toric CYs, for which powerful tools have been developed along the years.\footnote{Any CY singularity can be described as a real cone over a five-dimensional Sasaki-Einstein manifold. A toric CY is a CY for which the Sasaki-Einstein basis admits at least a $U(1)^3$ isometry group. Toric CYs can be obtained by partial resolutions of $\mathbf{C}^3/\textbf{Z}_n \times \textbf{Z}_m$, for large enough $m$ and $n$,  on which string theory can be consistently quantized, and so are open strings describing D-brane dynamics.} The corresponding effective gauge theories can be described in terms of quiver diagrams or more generally dimers, which encode all the information on the dual gauge theory (number of gauge groups, matter content and superpotential).

In \cite{Franco:2007ii} these tools were generalized to investigate gauge theories arising from D-branes at orientifolds of toric CYs. D-branes at orientifold singularities allow for a variety of new effects (novel kinds of gauge groups and matter representations, exotic instantons, new superpotential interactions, etc...), which can have diverse applications (see, {\it e.g.}, \cite{Blumenhagen:2006ci} for a review). In this paper, we would like to highlight a few properties of orientifold field theories which we believe have not been sufficiently emphasized so far, and which could have interesting implications.

Generically, there exist two classes of D-branes supported at a CY singularity. Regular branes, which can freely move in the full transverse space, and fractional branes, which can explore only a subspace of the transverse space. Fractional branes can be thought of as higher dimensional branes wrapping non-trivial (vanishing) cycles at the singularity and, as such, are (partially) stuck at the singularity (see \cite{Bertolini:2003iv} for a pedagogical review). 
Regular branes are known to support superconformal field theories while fractional branes lack some moduli and break conformal invariance. 

Orientifolds carry (negative) tension and RR charge and, when of high enough dimensionality, effective fractional brane charges as well. In the latter case, they break conformal invariance and contribute non-trivially to the effective dynamics of D-branes. This is the reason why gauge theories obtained by placing regular branes at orientifold singularities cannot be conformal by themselves (and, as we are going to see, this is also the reason why they end up being so interesting). Moreover, when adding fractional branes, orientifolds happen to interfere with them and, as we will discuss in detail,  change the expected UV and IR behaviors of the corresponding RG-flows. 

The breaking of conformal invariance due to the orientifold is a $1/N$ effect on the dynamics of $N$ regular branes. These are small effects at large $N$, but enough to make some gauge group factors become UV-free, together with others having instead vanishing $\beta$-function. This implies that the theory runs, irrespectively of the presence of explicit fractional brane sources. Moreover, while being $1/N$ suppressed, orientifold contributions get enhanced by the RG, and  can have dramatic effects on the dual gauge theory, enlarging the landscape of possible IR and UV behaviors one can describe placing D-branes at CY singularities. The main purpose of this paper is to discuss such effects. They can be summarized as follows:

\begin{enumerate}
\item One can consider bound states of regular and fractional branes at orientifold singularities and study the corresponding duality cascade. Unlike for unorientifolded singularities, the number of fractional branes is not preserved along the RG-flow. The most interesting consequence is that duality cascades are {\it finite}. Running the RG-flow up in energy, the theory reaches a free or nearly conformal phase after a finite RG-time. This implies that, unlike ordinary duality cascades which run indefinitely, orientifold cascades have UV-completions which do not require an infinite number of degrees of freedom.  
\item There exist quiver gauge theories which display runaway behavior in presence of fractional branes, as originally discussed in  \cite{Berenstein:2005xa,Franco:2005zu,Bertolini:2005di,Intriligator:2005aw}. Orientifolds typically cure this effect, stabilizing to supersymmetric vacua the runaway directions. 
\item Flavor branes can be added to orientifolds and, if properly chosen, they can compensate the orientifold tension and charge, recovering conformal invariance. The duality cascades in presence of such ``conformal" flavors become similar to ordinary ones. When adding a large number of flavor branes, one recovers the known pathology of a UV duality wall \cite{Benini:2007gx}.
\end{enumerate}

We will illustrate the above properties by discussing two specific models. However, the way these properties emerge suggests that they have a wider validity and hold for orientifold singularities in general.

The rest of the paper is organized as follows. In section \ref{chunf} we discuss our first example, an orientifold of  $\mathbf{F}_0$, the chiral $\mathbf{Z}_2$ orbifold of the conifold. This exhibits most of the properties summarized above, {\it i.e.}, the non-conservation of fractional brane charges and the finiteness of duality cascades. In section \ref{flccon} we consider the addition of flavor and show that, if properly chosen, they can restore conformal invariance. Perturbing such fixed points with fractional branes, one ends up with dynamics similar to ordinary cascades, including the possibility of having constant cascade steps, and, if the number of flavor is large enough, duality walls in the UV. In section \ref{secnonchi} we consider a different example, the orientifold of the non-chiral $\mathbf{Z}_2$ orbifold of the conifold. D-brane dynamics on this orientifold shares with the chiral one most of its properties. One interesting aspect here is that the parent orbifold theory allows for fractional brane configurations that lead to runaway. We show that the latter is cured by the orientifold via an effect which can be interpreted as a contribution of a stringy instanton at the bottom of the cascade or, following \cite{Aharony:2007pr}, as an effect generated one step up in the cascade and which propagates 
to low energy. Finally, section \ref{disc} contains a discussion on a few issues on which the peculiar properties of orientifold field theories might be relevant.

\section{The $\textbf{F}_0$ orientifold theory}
\label{chunf}

Let us focus on an orientifold line of  the $\textbf{F}_0$ theory, a chiral $\textbf{Z}_2$ orbifold of the conifold. We refer to \cite{Franco:2015kfa}, whose conventions we adopt, for details. 

\begin{figure}[ht]
\begin{center}
\includegraphics[height=0.22\textheight]{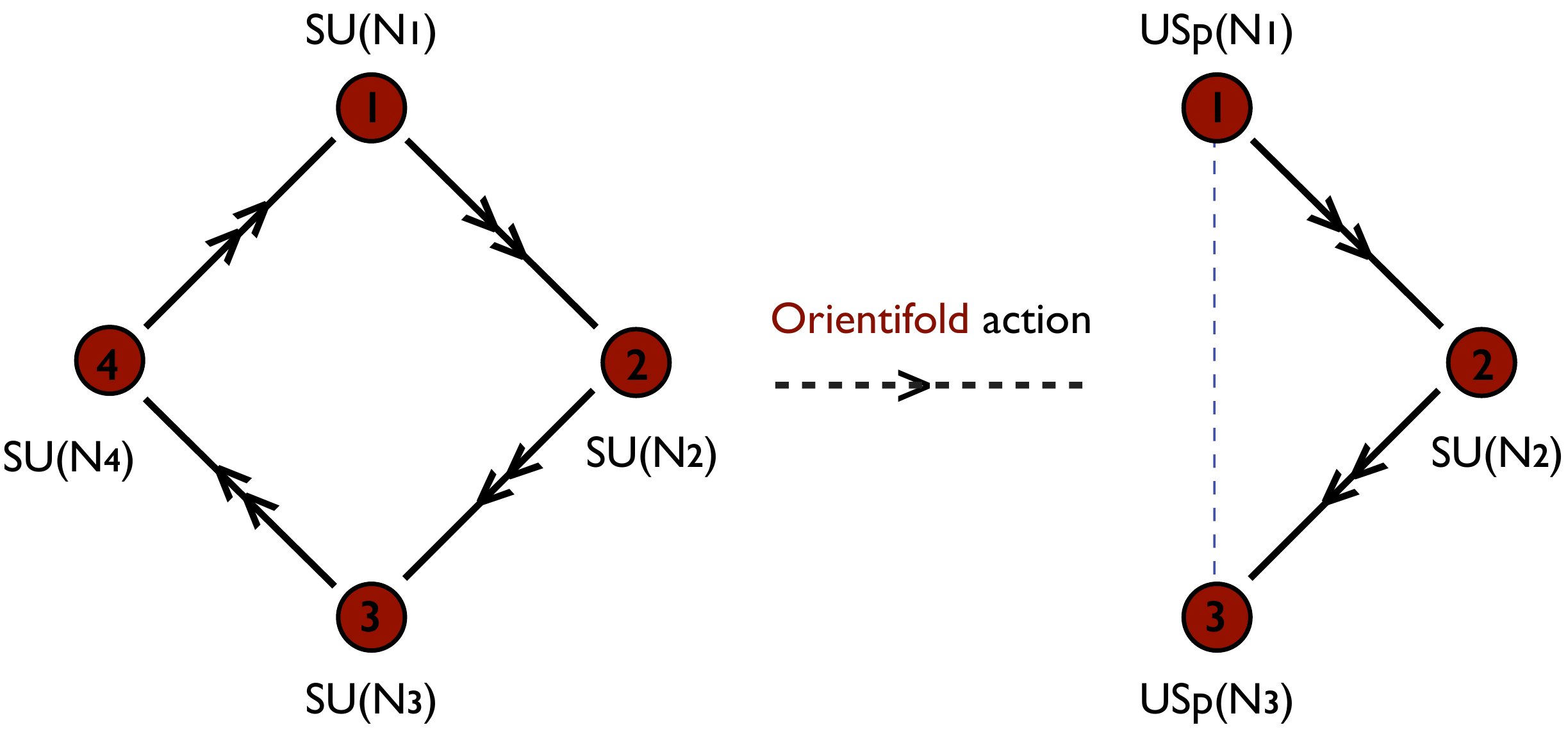}
\caption{Quiver diagram of $\textbf{F}_0$ on the left and of $\textbf{F}_0/\Omega$ on the right. The light blue line represents the fixed locus under the orientifold action and corresponds to a O7-plane. In our conventions $USp(N) = Sp(N/2)$, with $N$ even. In this way, at each node, be it $SU$ or $USp$, $N$ is the number of D-branes in the parent theory. Note that $N_1=N_3$ and $N_2=N_4$, in order to cancel gauge anomalies.}
\label{F0}
\end{center}
\end{figure}

The $\textbf{F}_0$ orientifold theory that we will consider is a $\cn=1$ supersymmetric gauge theory with three gauge groups, one unitary and two symplectic, chiral matter, a $SU(2)$ flavor symmetry and superpotential
\be
\label{sup1}
W= X_{12}^1X_{23}^1X_{23}^{2 \; T}X_{12}^{2 \; T} - X_{12}^1X_{23}^2X_{23}^{2 \; T}X_{12}^{1 \; T} -  X_{12}^2X_{23}^1X_{23}^{1 \; T}X_{12}^{2 \; T}~,
\ee
with obvious index notation. Quivers of the parent, non-orientifolded $\textbf{F}_0$ theory and that of the orientifold $\textbf{F}_0/\Omega$ are depicted in figure \ref{F0}.

Regular D3-branes at the singularity are described by quiver diagrams with $N=N_i$ at all nodes. The symmetries of the parent theory, the structure of the superpotential \eqref{sup1} and the relation between superconformal R-symmetry and operator dimensions, $\Delta = \frac 32 R$, fix the R-charges and scaling dimensions of all matter chiral superfields to be, respectively, $R=1/2$ and $\Delta=3/4$ (meaning that the anomalous dimension is $\gamma=-1/2$ for all matter fields). Given these quantum numbers, it is easy to check that for $N=N_i$ the parent theory has vanishing $\beta$-functions. On the contrary, the  orientifold theory $\beta$-functions get $1/N$ corrections due to the orientifold and read\footnote{The exact $\beta$-functions for unitarity and symplectic gauge groups are, respectively, $\beta(8\pi^2/g_{SU}^2) = 3 N_c - N_f (1-\gamma_f)$ and $\beta(8\pi^2/g_{USp}^2) = \frac32 (N_c +2) - \frac{N_f}2 (1-\gamma_f)$, with $\gamma_f$ the anomalous dimensions of matter fields. We take the latter to be $\gamma=-1/2$ also in the orientifold theory, even though there could be $1/N$ corrections to this value, too. It is easy to convince oneself that such corrections would not change qualitatively the physics. In particular, it would still be impossible to set all $\beta$-functions to zero.}
\be
\label{noMfr}
\beta^{USp} = 6 > 0\ , \qquad \qquad \beta^{SU}= 0\ ,
\ee
where here and in the following we denote by $\beta^{USp}$ the $\beta$-function for the sum of the two inverse square couplings of the $USp$ groups, taken to have coincident values. Eq.~\eqref{noMfr} means that only the $SU$ group enjoys a conformal phase while the two $USp$ groups are UV-free. Hence, differently from $\textbf{F}_0$, and more generally any non-orientifolded quiver gauge theory, the orientifold theory does not enjoy a stable conformal phase, once $1/N$ corrections are taken into account. This is due to the tension and RR charge carried by the O7-planes. Note that at a Calabi-Yau singularity, O7-planes, very much like D7-branes, do not only couple to the dilaton and the RR 8-form potential, but also to lower dimensional RR forms, and, hence, carry non-trivial fractional D3-brane charge \cite{Bertolini:2001qa}.
 
\subsection{The cascade}

In what follows, we would like to analyze the orientifold theory once fractional D3-branes are added to the system.\footnote{Duality cascades in presence of orientifolds have been discussed, among others, in \cite{Heckman:2007zp,Franco:2008jc,Amariti:2008xu,Franco:2010jv}.} The number of inequivalent fractional branes at a non-compact Calabi-Yau singularity equals the number of exceptional 2-cycles dual to non-compact 4-cycles. The $\textbf{F}_0$ singularity has one such 2-cycle and therefore admits one class of fractional branes, very much like the parent conifold theory. These branes are of the deformation type, {\it i.e.} they are associated to a complex structure deformation and can lead to confinement and gaugino condensation into isolated, supersymmetry preserving vacua.  

\begin{figure}[ht]
\begin{center}
\hskip -20pt\includegraphics[height=0.22\textheight]{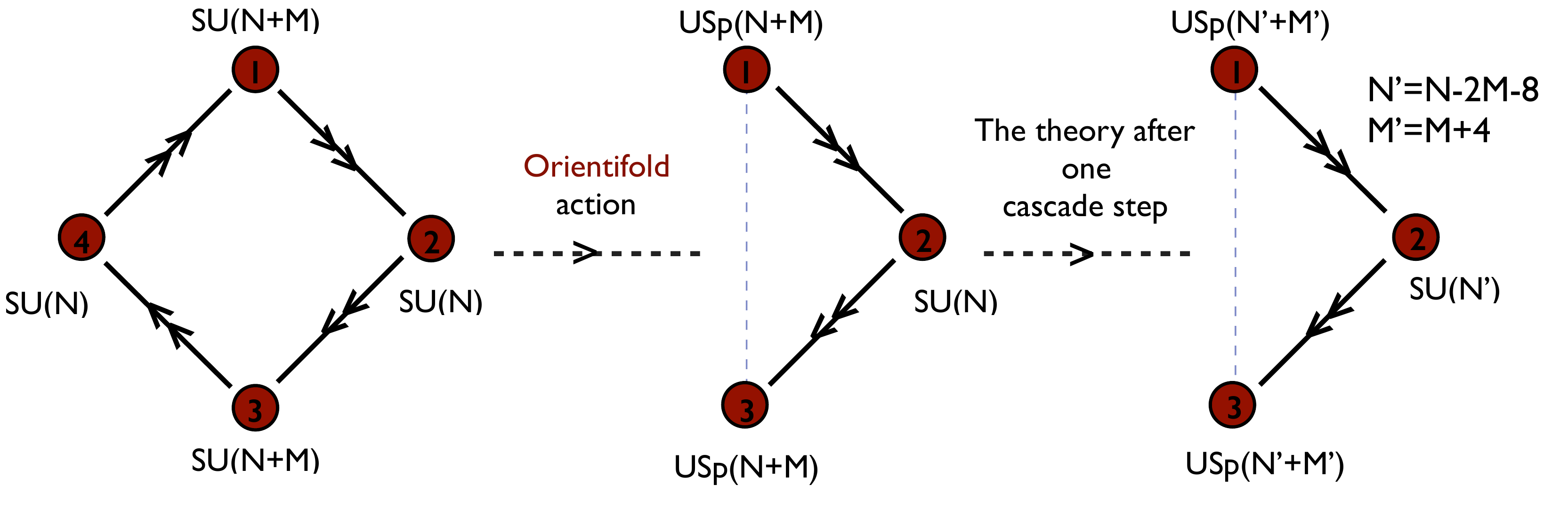}
\caption{$\textbf{F}_0/\Omega$ theory with fractional branes. On the left, the parent theory. On the extreme right, the orientifold theory after one cascade step.}
\label{F0fr}
\end{center}
\end{figure}

The addition of $M$ fractional branes on top of $N$ regular ones gives the extreme left quiver in figure \ref{F0fr}.  From the rank assignment point of view, $\textbf{F}_0$  can admit two types of fractional D3-branes: one affecting the rank of nodes 1 and 3, as shown in the figure, the other affecting the ranks of nodes 2 and 4. In fact, once taken together, they give rise to a regular brane. Since fractional branes are defined modulo regular branes, we choose, arbitrarily, a basis where fractional branes are those of the first type. As it will become clear in the next section, when discussing the addition of flavors, the O7-planes of the $\textbf{F}_0$ singularity, which carry fractional D-brane charge,  are associated to fractional branes of the first type.

Fractional branes induce a RG flow which can be interpreted as a cascade of Seiberg dualities. At each cascade step the theory returns to itself, with just a shift of the gauge group ranks. A cascade step corresponds to performing subsequent Seiberg dualities on nodes 1, 3, 2 and 4. This is inherited by the $\textbf{F}_0/\Omega$ theory, with the only caveat that nodes 2 and 4 are identified by the orientifold projection. As we discuss below, there are in fact further differences with respect to the parent $\textbf{F}_0$ singularity, which are subleading at large $N$ and $M$, but have relevant impact both on the UV and IR behaviors of the theory.

Let us analyze the duality cascades for $\mathbf{F}_0$ and  $\mathbf{F}_0/\Omega$, starting from a configuration with $N$ regular and $M$ fractional branes. 

In the parent non-orientifolded  theory, after one period the theory returns to itself, meaning that it is still described by the quiver on the left hand side of figure \ref{F0fr}, just with a shift in the number of regular branes $N \rightarrow N^\prime = N -2M$. 

The orientifold cascade is also self-similar, but $M$ also shifts. From the quiver in the center of figure \ref{F0fr}, one sees that the sympletic groups are those going to strong coupling first. Upon Seiberg duality, they change to $USp(N-M-4)$.\footnote{Seiberg duality acts differently on gauge group ranks for unitary and symplectic groups,  that is 
$SU(N_c)~,~ N_f \longrightarrow SU(N_f - N_c) ~,~N_f$ \cite{Seiberg:1994pq} while 
$USp(N_c)~,~N_f \longrightarrow  USp(N_f - N_c -4) ~,~N_f$ \cite{Intriligator:1995ne}.
For conciseness, in the following we will loosely call `rank'  the dimension of the fundamental representation of the group.} One then dualizes the $SU$ group, getting $SU(N-2M-8)$. The cascade is thus self-similar with $N'=N-2M-8$ and $M'=M+4$. It can also be checked, as in \cite{Franco:2015kfa}, that the superpotential \eqref{sup1} returns to itself after this sequence of Seiberg dualities. Hence, flowing towards the IR, the effective number of regular branes diminishes, while that of fractional branes increases. None of the two is invariant along the cascade, unlike in the parent theory, where $M$ stays constant along the flow. The non-conservation of $M$ along the RG flow is something familiar for ordinary quiver theories with flavors \cite{Benini:2007gx}. Quite opposite to those situations, though, 
in the present case the effective number of fractional branes becomes larger towards the IR. We will see that this fact has important implications.

Starting from a situation with, say, $N=N_0$ and $M=M_0$, after $k$ cascade steps the quiver returns the same with $N$ and $M$ given by
\begin{equation}
\label{cascnof}
N_k = N_0 - 2k (M_0+2k+2)\ , \qquad M_k = M_0 + 4k\  .
\end{equation}

The gauge group $\beta$-functions are easily computed, assuming that the anomalous dimensions of all bifundamental fields, fixed in the parent theory to be $\gamma=-1/2$, remain so all along the flow. Strictly speaking, this is true in the extreme UV, only, and becomes less and less so towards the IR, where $M$ and $N$ are of the same order and one should in principle take $M/N$ corrections into account. This is something shared by the parent conifold cascade and by duality cascades in general, where such corrections are known not to change, at least qualitatively, the RG-flow \cite{Strassler:2005qs}.\footnote{More precisely, one can argue that the particular RG flow trajectory which never gets close to the points in coupling space where one of the gauge groups is weakly coupled, is the one for which the anomalous dimensions stay very close to their conformal values. Interestingly, this is the flow best described by a smooth gravity dual, when it exists.} Finally, as we will discuss below, the IR dynamics, which is where such corrections would be most relevant, depends on arguments that do not rely on anomalous dimensions.    

At step $k$ the $\beta$-functions for the $USp$ and $SU$ nodes are, respectively, 
\begin{align}
\label{bu1}
\beta_{k}^{USp}&= 3(N_k+M_k+2)-3N_k = 3M_k+6\ , \\
\label{bu2}
\beta_{k}^{SU}&= 3N_k-3(N_k+M_k) =- 3M_k\ .
\end{align}
Relevant for our subsequent analysis are also the values of the $\beta$-functions at intermediate steps, namely after Seiberg duality on the $USp$ nodes has been performed, and that on the $SU$ node is yet to be done. They read 
\begin{align}
\label{bu12}
\beta_{k}^{USp}&= 3(N_k-M_k-2)-3N_k = -3M_k-6\ , \\
\label{bu22}
\beta_{k}^{SU}&= 3N_k-3(N_k-M_k-4) = 3M_k+12\ .
\end{align}

We would like now to consider the IR endpoint of the RG flow. In the parent $\textbf{F}_0$ theory if we start with $N= (2 l +1) M$, after $l$ cascade steps, nodes 1 and 3 have $N_f=N_c$, and enjoy a deformed moduli space, with mesonic and baryonic branches. As in the original conifold theory \cite{Klebanov:2000hb}, due to the quartic superpotential, the two branches are separated.  On the baryonic branch, the two $SU(2M)$ nodes confine and one is left with two decoupled pure SYM $SU(M)$ nodes, 2 and 4, which undergo gaugino condensation. 

In the orientifolded cascade things look similar, at first sight, since also in this case the theory can confine. However, some details are different and, as we will argue below, this has  important consequences. 

Let us first recall that, together with the number of regular branes, also that of fractional branes changes along the flow, now. In fact, their absolute value, cf.~eq.~\eqref{cascnof}, is not really physical. What really matters is the value of fractional branes modulo four.  
For the time being, let us assume, for simplicity, that $M_0=0$. In this case, after $k$ cascade steps the effective number of regular and fractional branes is 
\begin{equation}
N_k = N_0 - 4k (k+1)\ , \qquad M_k = 4k\ .
\end{equation}
In complete analogy with the non-orientifolded case, we aim at ending the cascade with the $SU$ node confining on the baryonic branch, followed by the $USp$ nodes becoming two decoupled pure SYM theories. For this to happen, we need the $SU$ node to have $N_f=N_c$ at the half-step. Supposing that this happens after $l$ steps (and a half), this fixes $N_0$ in terms of $l$ as
\begin{equation}
\left.
\begin{array}{l}
N_c= N_l = N_0-4l^2-4l \\
N_f= 2N_l -2M_l-8= 2(N_0-4l^2-8l-4)
\end{array} \right\rbrace  \Rightarrow  N_0 = 4l^2+12l +8\ .
\label{n0conf1}
\end{equation}
After the $SU$ group confines on the baryonic branch, we are left with two pure SYM $USp(N_l-M_l-4=4l+4)$ nodes, which undergo gaugino condensation and confine. If, in analogy with the conifold cascade, we call $M$ the rank of the pure SYM groups at the bottom of the cascade, we see that the number of steps in the cascade is a function of such rank 
\begin{equation}
\label{lMrel}
l=\frac{M-4}{4}\ ,
\end{equation}
as are the ranks of the quiver in the UV, when the cascade stops, see eq.~\eqref{noMfr}. They read
\begin{equation}
N_{UV}\equiv N_0 = \frac14 M^2 +M \ .
\end{equation}
Note that for $M\gg 1$ we have that $l\simeq M/4$ and $N_0 \simeq M^2/4$, {\it i.e.}, the number of steps is of the order of $M$, while the UV ranks are of order  $M^2$. 

We can also consider the case in which one ends up with pure SYM on the $SU$ group. For this to happen, we need, first, that the two $USp$ groups confine in a vacuum where all their mesons are massive. It is possible to see that, given the superpotential \eqref{sup1}, when the $USp$ nodes have (both) $N_f=N_c+2$ flavors, and thus confine on a deformed mesonic moduli space,  the latter cannot be completely removed. This is related to the fact that there does not exist a baryonic branch for $USp$ groups \cite{Intriligator:1995ne}. Luckily, one can see that taking the $USp$ groups to have $N_f=N_c+4$ flavors, the combination of the tree-level and dynamically generated superpotentials leads to an isolated confining vacuum for vanishing (and massive) mesonic operators. Let us recall how this comes about. Assuming we are left with pure $SU(M)$ SYM at the bottom of the cascade, the $USp$ groups must be $USp(2M-4)$. Calling the mesons of the two $USp$ groups (which are in the adjoint and anti-symmetric representations of $SU(M)$)
\begin{equation}
\mathcal{M}^{\alpha\beta}={X_{12}^{\alpha}}^TX_{12}^\beta\ , \qquad \qquad
\tilde{\mathcal{M}}^{\alpha\beta}=X_{23}^{\alpha}{X_{23}^\beta}^T\ ,
\end{equation}
the effective superpotential reads
\begin{equation}
W=\tilde{\mathcal{M}}^{12}\mathcal{M}^{21}-\tilde{\mathcal{M}}^{11}\mathcal{M}^{22}-\tilde{\mathcal{M}}^{22}\mathcal{M}^{11}+\mathrm{Pf}\mathcal{M}+\mathrm{Pf}\tilde{\mathcal{M}}\ ,
\end{equation}
where we have set all dimensionful factors to one to avoid clutter and $\mathcal{M}$ and $\tilde{\mathcal{M}}$ assemble all meson matrices $\mathcal{M}^{\alpha\beta}$ and $\tilde{\mathcal{M}}^{\alpha\beta}$, respectively. It is straightforward to show that for $\mathcal{M}=0=\tilde{\mathcal{M}}$ the F-term equations are satisfied, and all mesons are massive.

Let us then suppose that after $l$ cascade steps, we end up in a situation in which the $USp$ nodes have both $N_f=N_c+4$ flavors. This now fixes $N_0$ in terms of $l$ as
\begin{equation}
\left.
\begin{array}{l}
N_c= N_l +M_l= N_0-4l^2 \\
N_f= 2N_l = 2(N_0-4l^2-4l)
\end{array} \right\rbrace  \Rightarrow  N_0 = 4l^2+8l +4\ .
\end{equation}
After the $USp$ groups confine, we are left with pure SYM with gauge group $SU(N_l=4l+4)$. Hence, we end up again with the same relation \eqref{lMrel} between $l$ and $M$ while $N_0 = \frac14 M^2$. The large $M$ behavior of $l$ and $N_0$ is thus the same, irrespectively of which groups confine last. We will not distinguish anymore between these two cases. 

One can repeat the above analysis for all other inequivalent values of $M_0$, namely $M_0=1,2,3$, getting similar results, as far as the IR end of the cascade is concerned. More interesting is the UV behavior, where, as we discuss shortly, there is a sharp difference between $M_0$ being even or odd.

A common feature of the orientifold theory, regardless the value of $M_0$, is that, unlike the parent $\mathbf{F}_0$ theory, even in the absence of fractional branes it does not enjoy a conformal phase, and a duality cascade occurs. Since the effective number of regular branes diminishes along the RG-flow, $1/N$ corrections, which are negligible in the far UV, become more and more important towards the IR, eventually of order one. So important to let an otherwise conformal theory confine! 

Another important difference with the respect to the $\mathbf{F}_0$ theory regards the UV behavior. 

In the $\textbf{F}_0$ cascade (as in any ordinary non-orientifolded quiver gauge theory, as, {\it e.g.}, the conifold theory), the total rank, {\it i.e.}, the effective number of regular branes,  increases linearly and with no limits with the cascade steps as 
one goes to the UV. In this sense, the UV completion of such cascades needs an infinite number of degrees of freedom. 

\begin{figure}[ht]
\begin{center}
\includegraphics[height=0.22\textheight]{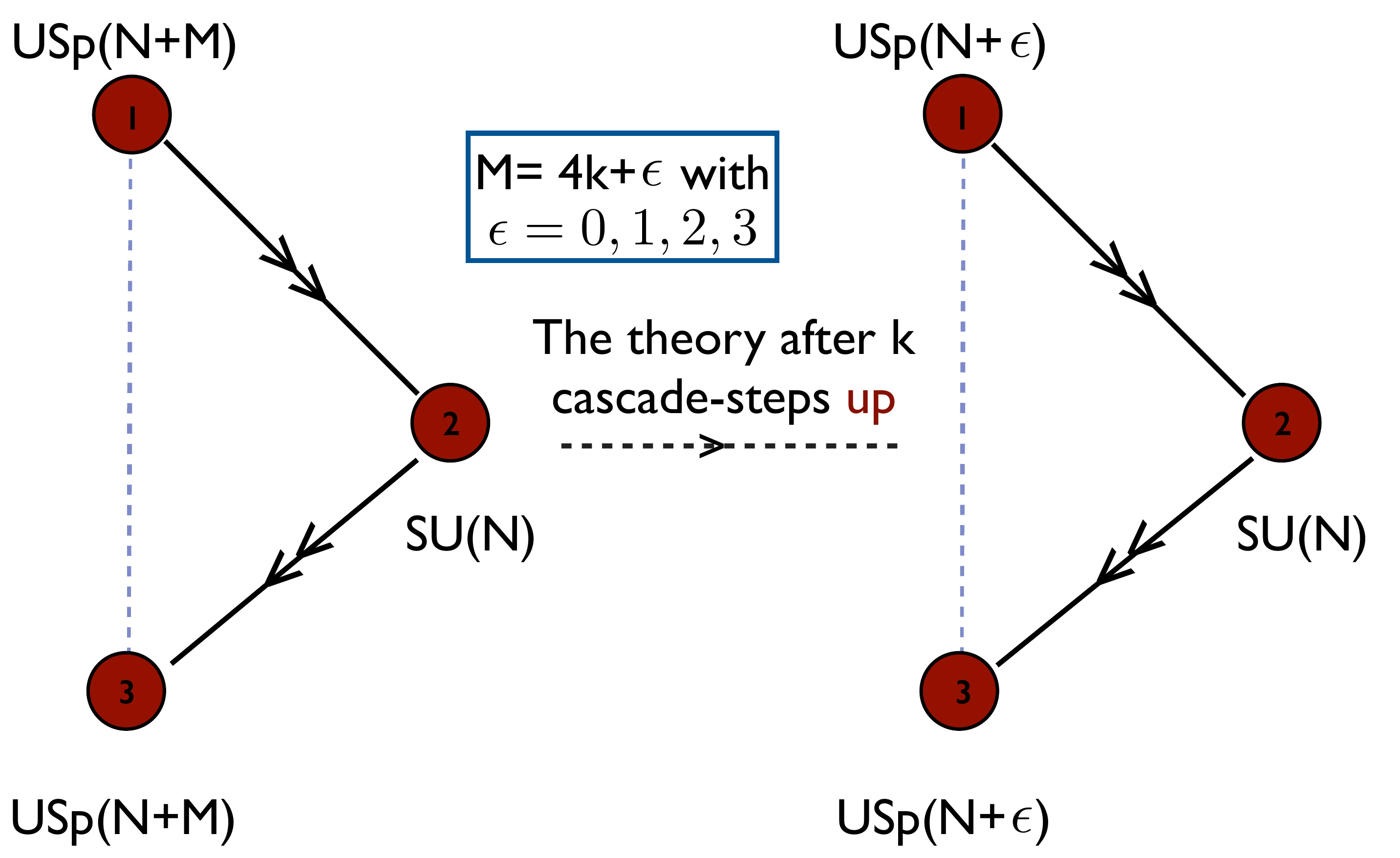}
\caption{The $\textbf{F}_0/\Omega$ theory with fractional branes. The inverse cascade. Different values of $\epsilon = 0,1,2,3$ define different UV completions. In the right quiver, we have renamed $N'$ as $N$.}
\label{F0u}
\end{center}
\end{figure}

For $\textbf{F}_0/\Omega$, the situation looks very different. 
Running the cascade up, the effective number of fractional branes diminishes, since $k$ does, see eq.~\eqref{cascnof}. So, in the far UV, it tends to zero. Depending on the value of $M$ (which, without loss of generality, we can parametrize as $M=4k+\e$ with $\e=0,1,2,3$), there are four different behaviors. Starting from some given values of $N$ and $M$, after $k$ steps up the quiver reduces to that on the right in figure \ref{F0u}.
Depending on the value of $\e$, the cascade stops or goes a few more steps up and then stops (this happens whenever there are no more IR free gauge groups). Below, we report the results for the four possible values of $\e$, after the very last cascade step has been performed
\begin{align}
\label{b1}
& \e=0  & \beta^{USp} = 6 ~~,~~\beta^{SU}= 0\ ,  \\
\label{b2}
& \e=1,3 &  \beta^{USp} = 3 ~~,~~\beta^{SU}= 3\ ,  \\
\label{b3}
& \e =2 &  \beta^{USp} = 0 ~~,~~\beta^{SU}= 6\ .  
\end{align}
We see that for $\e$ odd the UV-completions are free theories while for $\e$ even some groups reach an interacting fixed point. Note that the UV free nodes have a $\beta$-function which is subleading in $N$ (in other words, the $\beta$-function for the 't Hooft coupling is of order $1/N$).

A common feature of all four UV-completions, and the main qualitative difference with respect to ordinary cascades, is that the cascade stops at some large but {\it finite} rank. We are thus in presence of a cascade which consists in a finite number of steps, and as such a finite number of degrees of freedom. 

To summarize: the $\textbf{F}_0/\Omega$ theory does not have a conformal phase, independently on the number of regular and fractional branes. It enjoys a finite duality cascade and, in some circumstances, that is for $\e$ odd, the RG flow can interpolate between a genuine UV-free  theory down to a confining vacuum, like e.g. real-world QCD! 

In the coming subsection, we want to analyze the orientifold cascade a bit further. This will let us emphasize a few more interesting aspects of orientifold cascades which are not shared by standard ones, and will also prepare the ground for what we do next, namely the addition of dynamical  flavors.

\subsection{More details on the RG flow}

Let us discuss in more detail the RG evolution of the gauge couplings, in particular the length of the RG flow from the UV down to the deep IR. 

From eqs.~\eqref{b1}-\eqref{b3}, it is clear that there are just two qualitatively different flows, with odd or even $\epsilon$, respectively. They differ only at high energy since, as the energy diminishes, the effective number of fractional branes becomes large and order-one numbers would not matter much. In what follows, we will consider the case $\epsilon=0$ as prototype for the generic RG-flow, and then comment on the differences with the other flows.

Putting $\e=0$ into eqs.\eqref{bu1}-\eqref{bu22} we have 
\begin{equation}
\beta_{k}^{USp}= 12k+6\ , \qquad
\beta_{k}^{SU} =- 12k\ ,
\end{equation}
at any integral step, and 
\begin{equation}
\beta_{k}^{USp}= -12k-6\ , \qquad
\beta_{k}^{SU} =12k+12\ ,
\end{equation}
at any half-step of the duality cascade. We observe an increasing (absolute) value of the $\beta$-function as we go towards the IR, while their sum remains constant and equals six. Let us see how this translates in an estimate of the length of the RG flow. Recall that the RG equations are given by\footnote{We call here $1/ g^2_{USp}$ the sum of the two inverse squared couplings of the $USp$ groups.}
\begin{equation}
\frac{d}{dt}\frac{1}{g^2_{USp}}=\beta^{USp}\ , \qquad 
\frac{d}{dt}\frac{1}{g^2_{SU}}=\beta^{SU}\ ,
\end{equation}
where $t=\frac{1}{8\pi^2}\log \mu$ and $\mu$ is the RG scale. 
At the integral step $k$ we can assume that 
\begin{equation}
\frac{1}{g^2_{SU}(t_k)}=0\ , 
\end{equation}
and define
\begin{equation}
\frac{1}{g^2_{USp}(t_k)}=\frac{1}{g_k^2}\ .
\end{equation}
The RG evolution of the couplings is then
\begin{equation}
\label{RG11}
\frac{1}{g^2_{USp}(t)}=(12k+6)(t-t_k) + \frac{1}{g_k^2}\ , \qquad
\frac{1}{g^2_{SU}(t)}=-12k(t-t_k)\ .
\end{equation}
We need to perform a Seiberg duality on the $USp$ nodes when their couplings diverge. This occurs at the RG time $t'_k$ such that 
\begin{equation}
\frac{1}{g^2_{USp}(t'_k)}=0 \qquad \Leftrightarrow \qquad t'_k-t_k= - \frac{1}{(12k+6)g_k^2}\ .
\end{equation}
At $t=t'_k$ the $SU$ coupling is instead 
\begin{equation}
\frac{1}{{g'_k}^2}=\frac{1}{g^2_{SU}(t'_k)}=-12k(t'_k-t_k)=\frac{12k}{(12k+6)g_k^2}\ .
\end{equation}
We now continue the RG evolution with the half-step $\beta$-functions. At the RG time $t_{k+1}$ the $SU$ node becomes, again, infinitely strongly coupled. This is determined to be
\begin{equation}
t_{k+1}=t'_k -\frac{1}{(12k+12){g'_k}^2}=t_k-\frac{1}{6(k+1)g_k^2}\ ,
\end{equation}
while the coupling of the $USp$ nodes is
\begin{equation}
\frac{1}{g_{k+1}^2}= -(12k+6)(t_{k+1}-t'_k)=\frac{k}{(k+1)g_k^2}=\frac{1}{(k+1)g_1^2}\ ,
\end{equation}
where in the last equality we have used the expression recursively. We see that for a long enough flow, {\it i.e.}, for $k$ sufficiently large, we have that 
\begin{equation}
\label{rune0}
\frac{1}{g_{k}^2}= \frac{1}{k g_1^2}\ , \qquad 
t_{k+1}-t_k \simeq -\frac{1}{6k^2 g_1^2}\ .
\end{equation}
This implies that the cascade steps become more and more dense towards the IR, while the sum of the inverse squared couplings is controlled by the sum of the $\beta$-functions, which is 
\begin{equation}
\beta_{k}^{USp}+\beta_{k}^{SU}= 6 
\end{equation}
all along the flow, independently of $k$. Note that this is an order $1/N$ effect. It implies that the envelope within which the $1/g_k^2$ are evolving is tilted downwards towards the IR, in agreement with the fact that we have (at least) an asymptotically free gauge group in the UV. 

It is easy to see that the cascade stops before the envelope crosses the horizontal axis, where a (IR) duality wall would occur. From eqs.~\eqref{RG11} one sees that the crossing happens at $t_\mathrm{wall}=-1/6g_1^2$, if we set $t_1=0$. 
Assuming $l$ large, the length of the flow stopping at $t_l$ reads\footnote{We have used the fact that $\sigma_n=\sum_{k=1}^n\frac{1}{k(k+1)}=\frac{n}{n+1}$, which can be verified noticing that $\sigma_n-\sigma_{n-1}=\frac{n}{n+1}-\frac{n-1}{n}=\frac{1}{n(n+1)}$.} 
\begin{equation}
t_l = \sum_{k=1}^{l-1} (t_{k+1}- t_{k})=-\frac{1}{6g_1^2}\sum_{k=1}^{l-1}\frac{1}{k(k+1)}= -\frac{1}{6g_1^2}\left(1-\frac{1}{l}\right)\gtrsim t_\mathrm{wall}\ .
\end{equation}
 Hence, we see that confinement happens very close in energy to the accumulation point (or duality wall), but yet an infinite number of dualities away. The IR scale $\Lambda_l\simeq \Lambda_\mathrm{wall}$ is related to the UV scale $\Lambda_0$ as
\begin{equation}
\mbox{$\textbf{F}_0/\Omega$:}\qquad\qquad \Lambda_l=\Lambda_0 e^{-\frac{1}{6g_1^2} }\ .
\end{equation}
This is to be confronted with, {\it e.g.}, the conifold cascade, where we have
\begin{equation}
\mbox{Conifold:}\qquad\qquad \Lambda_l=\Lambda_0 e^{-\frac{2}{3Mg^2} l}\ ,
\end{equation}
with $M$ the rank of the SYM gauge group at the bottom of the cascade. In both theories $N_0\simeq l M$, and in $\textbf{F}_0/\Omega$ we have also that $N_0\simeq M^2/4$, so we  can rewrite the two hierarchies in the following suggestive way
\begin{equation}
\mbox{Conifold:}\qquad \Lambda_l=\Lambda_0 e^{-\frac{2N_0}{3M^2g^2} }\ ,\qquad\qquad\qquad \mbox{$\textbf{F}_0/\Omega$:}\qquad \Lambda_l=\Lambda_0 e^{-\frac{2N_0}{3M^2g^2} }\ .
\end{equation}
We are thus in presence of a similar hierarchy in the orientifold theory, with the notable difference that here one can follow the whole RG flow, from the UV to the IR (and not from an arbitrary scale, as in the conifold). This is due to the UV-completion of the orientifold cascade which, as already emphasized, does not require an infinite number of degrees of freedom.
\begin{figure}[t]
\begin{center}
\includegraphics[height=0.42\textheight]{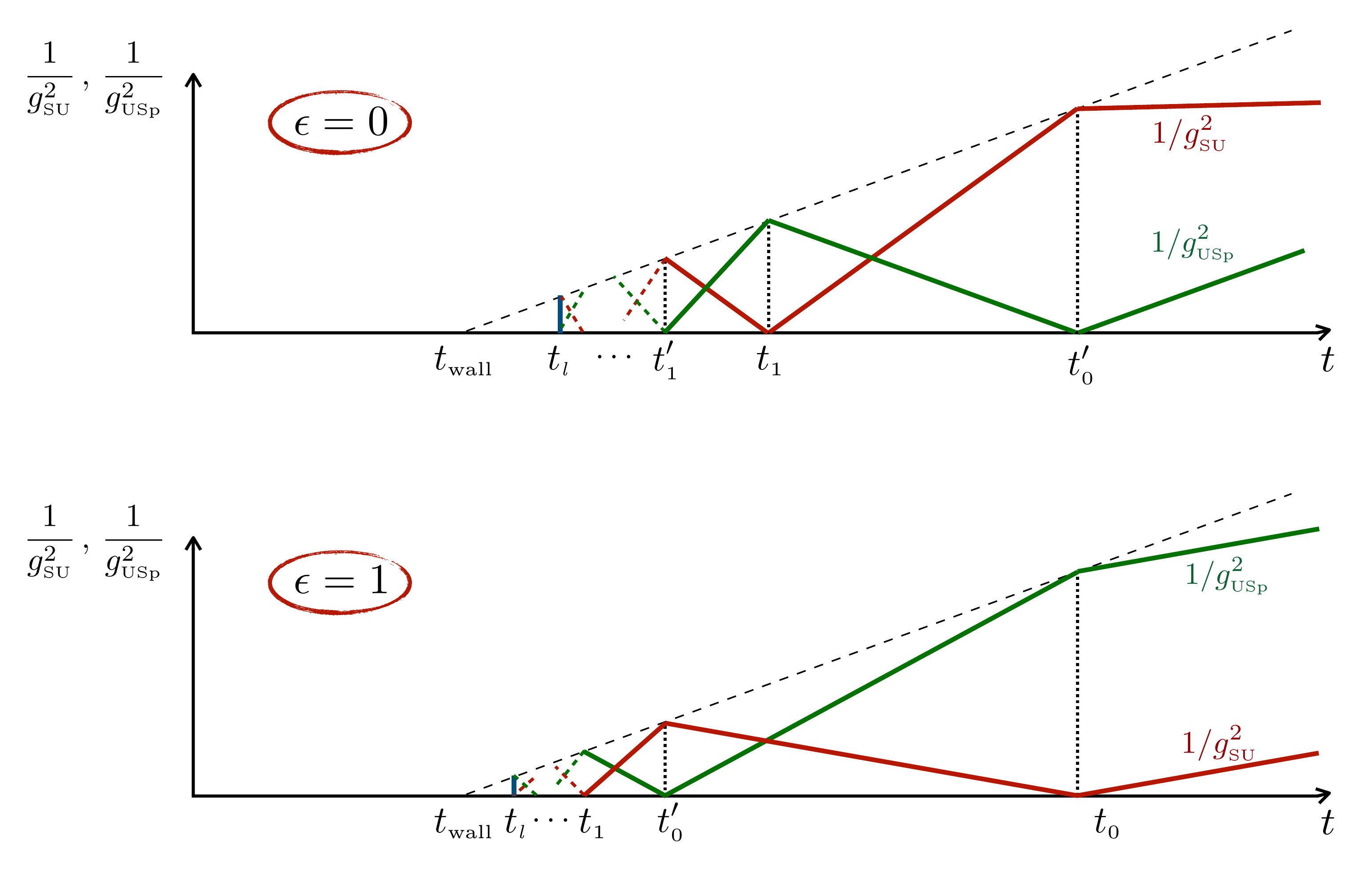}
\caption{The different behaviors of the gauge coupling running for $\e$ even and odd, respectively (as prototypes, only $\e=0,1$ are drawn). After the very last step {\it up} has been performed, for $\e$ odd all couplings run the same and are UV-free. For $\e$ even, one of the couplings stops running, instead (the $SU$ coupling for $\e=0$ and the $USp$ coupling for $\e=2$). The IR behaviors are instead similar. In particular, confinement occurs at $t=t_l$ and is always reached before hitting the (would-be) duality wall.}
\label{runUV}
\end{center}
\end{figure}

One can repeat the above analysis for any other allowed values of $\e$. As far as the IR end of the cascade, one would get qualitative similar answers. As far as the UV, instead, even and odd $\e$ behave differently. Starting from the IR and running the cascade upwards, at each duality step the $USp$ and $SU$ $\beta$-functions have opposite signs, up to $k=0$, where $M=\e$. At that point, one should perform one more step up and ends up with eqs.\eqref{b1}-\eqref{b3}. The main difference is that for $\e=0,2$ some gauge couplings become free, and some other reach a fixed point and stop running. For $\e$ odd, instead, all gauge couplings become free, and so one reaches a UV-free trivial fixed point. These differences are depicted in figure \ref{runUV}.

In all the analysis above, we have neglected $M/N$ corrections to matter field anomalous dimensions, which are expected to become more and more important towards the IR. These effects deform the cascade \cite{Strassler:2005qs} and might then modify, {\it e.g.}, the distance between subsequent cascade steps $\Delta t_k = |t_{k} - t_{k-1}|$ for $k$ large and close to $l$. We do not expect this to change the basic property of cascade steps becoming the more and more dense the larger $k$, {\it i.e.} $\Delta t_{k} < \Delta t_{k-1}$, which is at the basis of the confinement before the wall behavior we have found. A concrete way to check this prediction would be to find the gravity dual description of these flows, which, as already noted, corresponds to a trajectory in coupling space where such effects are more under control. Work in this direction is under way \cite{RM2}.

An important comment is worth at this point. Until now, we have only considered the case of an orientifold projection leading to $USp$ groups. One could ask what would have changed had we considered the opposite projection, leading to $SO$ groups instead. Technically, the main difference is that, due to the flipping of some signs, the number of fractional branes would increase towards the UV, now. This has quite dramatic consequences. Basically, the RG flow would be the mirror image of the one in figure \ref{runUV}: the envelope would be tilted upwards towards the IR now, and a duality wall would show-up in the UV. This is a behavior already observed in \cite{Benini:2007gx} in the flavored conifold theory (and, as we will discuss in  section \ref{flccon}, emerging also in our set-up, if a large enough number of flavors is added). We thus conclude that O7-planes leading to $USp$ groups are singled out in leading to the very peculiar behavior of finite cascades.

\subsubsection*{Metastable supersymmetry breaking vacua}

In the conifold theory, if the number of regular branes, $N$, is not taken to be proportional to that of fractional branes, $M$, the baryonic branch at the bottom of the cascade is lifted \cite{Dymarsky:2005xt}. In  
\cite{Kachru:2002gs} it was argued that on the would-be baryonic branch there exist local minima of the potential (which, as such, are metastable), described by antiD3-branes at the tip of the deformed conifold, which embiggen into an NS5 shell. These vacua have the same quantum numbers of the (still existing) mesonic branch, to which they decay in finite (but, under certain conditions, parametrically large) time. 

The situation for the $\mathbf{F}_0/\Omega$ theory looks quite similar. Here, too, the baryonic branch of, say, the $SU$ node is lifted as soon as one does not choose the number of regular branes to satisfy eq.~\eqref{n0conf1} (or its siblings for $\epsilon=1,2,3$). The orbifolding procedure leading to $\mathbf{F}_0$ and the subsequent orientifolding leading to  $\mathbf{F}_0/\Omega$ do not appear to change the basic dynamics leading to local minima of the potential. Hence, one can expect that also in the present case metastable supersymmetry breaking vacua might exist.\footnote{It is worth mentioning that in the non-chiral $\mathbf{Z}_2$ orbifold of the conifold, and the corresponding orientifold, which we will discuss in section \ref{secnonchi}, this has been shown to be the case \cite{Argurio:2006ny,Argurio:2007qk}.}  A notable difference is that here the cascade is finite and the UV is essentially conformal (at leading order, and asymptotically free at  subleading order). Hence, these metastable states created by anti-branes on orientifolded deformed singularities should be under better control with respect to non-orientifolded ones.

\section{The flavored $\textbf{F}_0$ orientifold}
\label{flccon}

The dynamics described in the previous section can sensibly change when adding flavors. Two peculiar properties of the unflavored orientifold is that there do not exist conformal phases and that cascade steps are not constant, in the sense that the effective number of fractional branes changes along the cascade (together with that of regular branes). These properties are both due, eventually, to the presence of the O7-planes. One could ask if adding flavor D7-branes, which can possibly compensate the O-plane tension and charge, this state of affairs can change and, also, if the corresponding cascades, when fractional branes are added, can enjoy different  dynamics. As we are going to show below, the answers to both such questions is positive.

We will start addressing the question on whether there exist D-brane configurations on $\textbf{F}_0/\Omega$ describing (interacting) SCFTs.\footnote{A similar question, albeit for non-cascading orientifold field theories, was addressed in, {\it e.g.}, \cite{Bianchi:2013gka}.} Later, we will consider arbitrary numbers of flavors, as well as the addition of fractional branes, and we will discuss the corresponding RG flows.

\subsection{Flavors and superconformal phases}
\label{flconf}

There are two alternative ways to add flavors to the conifold field theory, related to two different supersymmetric embeddings of D7-branes, the so-called Kuperstein embedding \cite{Kuperstein:2004hy} and Ouyang embedding \cite{Ouyang:2003df}. These two different possibilities are inherited by the $\textbf{F}_0/\Omega$ geometry. If we aim at recovering conformal invariance we need to add D7-branes so to cancel exactly the O7 tension and charges. As we are going to show below, this is obtained following the Kuperstein embedding for the flavor branes (which shows that the O-planes enjoy that same embedding).  

\begin{figure}[ht]
\begin{center}
\includegraphics[height=0.26\textheight]{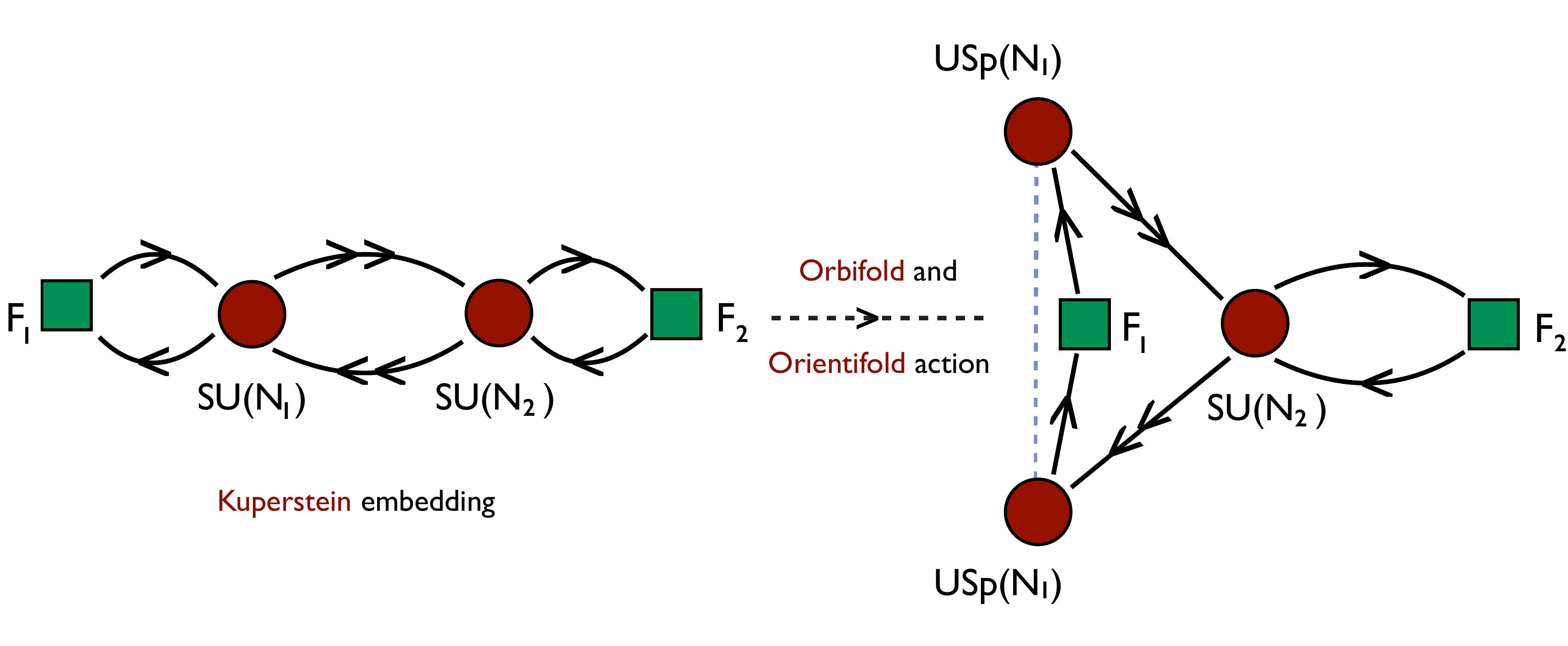}
\caption{The conifold and the $\textbf{F}_0/\Omega$ theories with (Kuperstein-like) flavors.}
\label{FQ}
\end{center}
\end{figure}

In figure \ref{FQ}, we report the quivers of the flavored theories. The left quiver represents the flavored conifold theory while the right quiver the flavored orientifold theory.

As we see from the figure, the flavored $\textbf{F}_0/\Omega$ quiver is given by the addition of the following fields in the (anti)fundamental representation: $F_1$ fundamentals $Q_1$ for the $USp_1$ gauge group, the same number $F_1$ of fundamentals $Q_3$ for the $USp_3$ gauge group, and $F_2$ pairs of fundamentals $Q_2$ and anti-fundamentals $\tilde Q_2$ for the $SU$ group. The fact that the number of flavors for the two $USp$ groups is the same is inherited from the parent conifold and $\textbf{F}_0$ theories, by requiring the absence of gauge anomalies.

The full superpotential reads
\begin{align}
\label{supflav}
W= &\ X_{12}^1X_{23}^1X_{23}^{2 \; T}X_{12}^{2 \; T} - X_{12}^1X_{23}^2X_{23}^{2 \; T}X_{12}^{1 \; T} -  X_{12}^2X_{23}^1X_{23}^{1 \; T}X_{12}^{2 \; T} \nonumber \\
& +Q_1 X_{12}^1 X_{23}^1 Q_3 +Q_1 X_{12}^2 X_{23}^2 Q_3 + Q_2 X_{23}^1 {X_{23}^1}^T Q_2^T \nn\\
&+Q_2 X_{23}^2 {X_{23}^2}^T Q_2^T +\tilde{Q}_2^T {X_{12}^1}^T X_{12}^1 \tilde{Q}_2 +
\tilde{Q}_2^T {X_{12}^2}^T X_{12}^2 \tilde{Q}_2 \nn \\
&+\frac12 Q_1 Q_1^T Q_3 Q_3^T - Q_2\tilde Q_2 Q_2\tilde Q_2 \ .
\end{align}
All terms in the above superpotential are quartic and, based on symmetry arguments, one can see that all matter fields, including the $Q$'s, have R-charge $1/2$ and $\gamma=-1/2$. The $\beta$-functions read
\begin{align}
\beta^{USp}&= 3(N_1+2)-3N_2-\frac32 F_1 \ , \\
\beta^{SU}&= 3N_2-3N_1-\frac32 F_2 \ ,
\end{align}
and vanish simultaneously if
\begin{equation}
\label{scfrel}
F_1+F_2=4\ , \qquad \qquad N_2-N_1=\frac12 F_2\ .
\end{equation}
We have thus 3 possible superconformal field theories
\begin{align}
F_1 = 4 \ , \qquad & F_2= 0 \ , \qquad  N_2= N_1 \ , \\
F_1 = 2 \ , \qquad & F_2= 2 \ , \qquad  N_2= N_1 +1\ , \\
F_1 = 0 \ , \qquad & F_2= 4 \ , \qquad  N_2= N_1+2 \ .
\end{align}
It is obvious that the first case, with only 4 $F_1$-type flavors, is the one corresponding to the D7-branes lying exactly on the O7. In the two other cases, some or all of the D7s are still parallel to the O7, but, in the language of the parent theory, with fractional D-brane charge of the second type, hence leading to the small mismatch in the ranks of the $USp$ and $SU$ gauge groups.\footnote{This interpretation can also be checked looking at the embeddings of the $F_1$ and $F_2$ flavor branes in the original conifold geometry. The conifold can be described as an hypersurface in $\mathbf{C}^4$, $xy-wz=0$. The D7-brane embedding \cite{Kuperstein:2004hy} is described by the equation $w+z=0$, which is invariant under the $\mathbf{Z}_2$ orbifold action $ x \rightarrow -x , y \rightarrow - y ,  w \rightarrow -w  , z \rightarrow -z $, and defines the O7-planes, too. The $F_1$ flavors correspond to D7s with no gauge flux on their worldvolume, while $F_2$ flavors corresponds to D7s with worlvolume flux \cite{Benini:2007gx}.}

Consistently with conformal invariance, one can easily check that if relations \eqref{scfrel} are satisfied, not only the $\beta$-functions vanish, but the quiver is also self-dual under Seiberg duality. 

As already mentioned, one could consider a different kind of flavors, related to the so-called Ouyang embedding. This would give rise to flavors which intervene in cubic couplings in the superpotential. It is easy to see that in this case it is impossible to achieve a configuration of gauge and flavor ranks such that the theory is both conformal and self-similar under Seiberg dualities, meaning that such flavors cannot fully cancel the orientifold charges, as anticipated. We will not consider this case further.

Besides leaving the gauge group ranks unchanged, one should also show that under a sequence of Seiberg dualities the superpotential  \eqref{supflav} is self-similar. That this is the case will be shown below in full generality, namely regardless the values of $N_1, N_2, F_1$ and $F_2$. 

\subsubsection*{Self-similarity of the (flavored) superpotential}
Let us consider the flavored orientifold quiver in figure \ref{FQ}, with arbitrary values of  $N_1, N_2, F_1$ and $F_2$.

Dualizing the $USp$ nodes, we introduce the following mesons
\begin{equation}
\label{mes1}
\mathcal{M}_{22}^{\alpha\beta}= {X_{12}^\alpha}^TX_{12}^\beta\ , \qquad \mathcal{N}_2^\alpha = Q_1X_{12}^\alpha\ , \qquad \mathcal{P}=Q_1Q_1^T\ ,
\end{equation}
\begin{equation}
\label{mes2}
\tilde{\mathcal{M}}_{22}^{\alpha\beta}= X_{23}^\alpha{X_{23}^\beta}^T\ , \qquad \tilde{\mathcal{N}}_2^\alpha = X_{23}^\alpha Q_3\ , \qquad \tilde{\mathcal{P}}=Q_3Q_3^T\ ,
\end{equation}
and the dual fields
\begin{equation}
Y_{21}^\alpha \ , \qquad Y_{32}^\alpha \ ,  \qquad q_1\ , \qquad q_3\ .
\end{equation}
The superpotential becomes
\begin{align}
W= &\ \mathcal{M}_{22}^{21} \tilde{\mathcal{M}}_{22}^{12}  -\mathcal{M}_{22}^{11} \tilde{\mathcal{M}}_{22}^{22}
-\mathcal{M}_{22}^{22} \tilde{\mathcal{M}}_{22}^{11} + \mathcal{N}_2^1  \tilde{\mathcal{N}}_2^1 + \mathcal{N}_2^2  \tilde{\mathcal{N}}_2^2+\frac12\mathcal{P} \tilde{\mathcal{P}} 
\nonumber \\
& + Q_2 \tilde{\mathcal{M}}_{22}^{11} Q_2^T+Q_2  \tilde{\mathcal{M}}_{22}^{22} Q_2^T +\tilde{Q}_2^T \mathcal{M}_{22}^{11}\tilde{Q}_2 +
\tilde{Q}_2^T \mathcal{M}_{22}^{22} \tilde{Q}_2 - Q_2\tilde Q_2 Q_2\tilde Q_2  \nn \\
& + \mathcal{M}_{22}^{21} Y_{21}^1 {Y_{21}^2}^T + \mathcal{M}_{22}^{11} Y_{21}^2 {Y_{21}^2}^T  + \mathcal{M}_{22}^{22} Y_{21}^1 {Y_{21}^1}^T  + \frac12\mathcal{P}q_1q_1^T\nn\\
&+  \tilde{\mathcal{M}}_{22}^{12}{Y_{32}^2}^T Y_{32}^1+ \tilde{\mathcal{M}}_{22}^{11}{Y_{32}^2}^T Y_{32}^2  + \tilde{\mathcal{M}}_{22}^{22}{Y_{32}^1}^T Y_{32}^1 +\frac12\tilde{\mathcal{P}}q_3q_3^T \nn\\
& + \mathcal{N}_2^1 Y_{21}^1 q_1 + \mathcal{N}_2^2 Y_{21}^2 q_1 +  q_3{Y_{32}^1}\tilde{\mathcal{N}}_2^1 
+ q_3{Y_{32}^2}\tilde{\mathcal{N}}_2^2\ .
\end{align}
Because of the first line, all mesons are massive and can be integrated out, yielding
\begin{align}
W= &\  -Y_{32}^1Y_{21}^1 {Y_{21}^2}^T{Y_{32}^2}^T +Y_{32}^1Y_{21}^2 {Y_{21}^2}^T{Y_{32}^1}^T+Y_{32}^2Y_{21}^1 {Y_{21}^1}^T{Y_{32}^2}^T
\nonumber \\
&  +Q_2 Y_{21}^1 {Y_{21}^1}^T Q_2^T+Q_2 Y_{21}^2 {Y_{21}^2}^T Q_2^T
+\tilde{Q}_2^T {Y_{32}^1}^T Y_{32}^1 \tilde{Q}_2+ \tilde{Q}_2^T {Y_{32}^2}^T Y_{32}^2 \tilde{Q}_2 \nn \\
& - q_3 Y_{32}^1 Y_{21}^1 q_1 - q_3 Y_{32}^2 Y_{21}^2 q_1+ Q_2\tilde Q_2 Q_2\tilde Q_2 -\frac12q_1q_1^Tq_3q_3^T \ .
\end{align}
This superpotential is already self-similar to \eqref{supflav}, up to some signs and charge conjugations.  We now perform Seiberg duality on the $SU$ node. The mesons are
\begin{equation}
\mathcal{M}_{31}^{\alpha\beta}= Y_{32}^\alpha Y_{21}^\beta\ , \qquad \mathcal{N}_1^\alpha = Q_2Y_{21}^\alpha\ , \qquad \mathcal{N}_3^\alpha = Y_{32}^\alpha \tilde{Q}_2\ ,\qquad \mathcal{R}=Q_2\tilde{Q}_2\ ,
\end{equation}
and the dual fields
\begin{equation}
Z_{12}^\alpha \ , \qquad Z_{23}^\alpha \ ,  \qquad q_2\ , \qquad \tilde{q}_2\ .
\end{equation}
The superpotential reads
\begin{align}
W= &\  -\mathcal{M}_{31}^{11} {\mathcal{M}_{31}^{22}}^T +\mathcal{M}_{31}^{12} {\mathcal{M}_{31}^{12}}^T +\mathcal{M}_{31}^{21} {\mathcal{M}_{31}^{21}}^T +\mathcal{N}_1^1{\mathcal{N}_1^1}^T +\mathcal{N}_1^2{\mathcal{N}_1^2}^T  \nn \\
& +{\mathcal{N}_3^1}^T\mathcal{N}_3^1+{\mathcal{N}_3^2}^T\mathcal{N}_3^2+\mathcal{R}^2- q_3 \mathcal{M}_{31}^{11} q_1 - q_3 \mathcal{M}_{31}^{22} q_1 -\frac12q_1q_1^Tq_3q_3^T  \nn \\
& + \mathcal{M}_{31}^{11} Z_{12}^2 Z_{23}^2 +  \mathcal{M}_{31}^{22} Z_{12}^1 Z_{23}^1+ 2\mathcal{M}_{31}^{12} Z_{12}^2 Z_{23}^1+ 2\mathcal{M}_{31}^{21} Z_{12}^1 Z_{23}^2\nn \\
&+2q_2 Z_{23}^1 \mathcal{N}_3^1 +2q_2 Z_{23}^2 \mathcal{N}_3^2 +2 \mathcal{N}_1^1 Z_{12}^1 \tilde{q}_2 +2 \mathcal{N}_1^2 Z_{12}^2 \tilde{q}_2 + 2\mathcal{R}q_2\tilde{q}_2\ ,
\end{align}
where the numerical factors in the last two lines have been put for convenience.
Again, all mesons are massive and can be integrated out, yielding 
\begin{align}
W= &\ Z_{12}^1Z_{23}^1Z_{23}^{2 \; T}Z_{12}^{2 \; T} - Z_{12}^1Z_{23}^2Z_{23}^{2 \; T}Z_{12}^{1 \; T} -  Z_{12}^2Z_{23}^1Z_{23}^{1 \; T}Z_{12}^{2 \; T} \nonumber \\
& -q_1^T Z_{12}^1 Z_{23}^1 q_3^T -q_1^T Z_{12}^2 Z_{23}^2 q_3^T - q_2 Z_{23}^1 {Z_{23}^1}^T q_2^T \nn\\
&-q_2 Z_{23}^2 {Z_{23}^2}^T q_2^T -\tilde{q}_2^T {Z_{12}^1}^T Z_{12}^1 \tilde{q}_2 -
\tilde{q}_2^T {Z_{12}^2}^T Z_{12}^2 \tilde{q}_2 \nn \\
&+\frac12q_1 q_1^T q_3 q_3^T - q_2\tilde q_2 q_2\tilde q_2 \ .
\end{align}
This is exactly the same as \eqref{supflav}, up to a trivial relabeling of the fields $Z_{ij}^\alpha \to X_{ij}^\alpha$, 
$q_{1,3}^T \to i Q_{1,3}$ and $q_2, \tilde{q}_2 \to iQ_2, i\tilde{Q}_2$. This shows that the superpotential \eqref{supflav} is indeed self-similar under a cascade step, for generic values of $N_1, N_2, F_1$ and $F_2$.

\subsection{Back to normal: cascading with flavors}

Having checked that the superpotential is self-similar under a cycle of Seiberg dualities, we can now consider the most general rank assignment, and study the corresponding RG flow.

\begin{figure}[ht]
\begin{center}
\includegraphics[height=0.28\textheight]{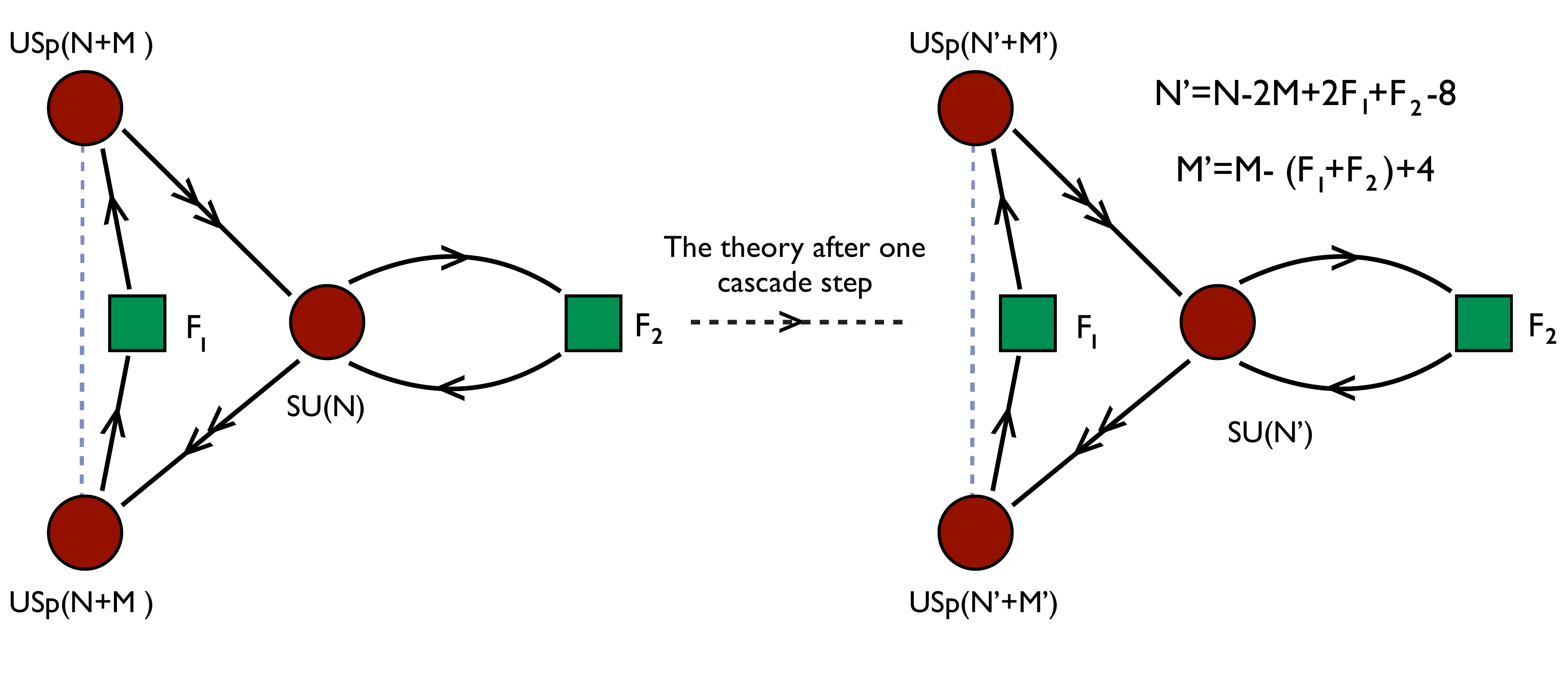}
\caption{The effect of a cascade step on the $\textbf{F}_0/\Omega$ theory with flavors.}
\label{FQcas}
\end{center}
\end{figure}

To be concrete, let us consider $USp(N_1=N+M)$ and $SU(N_2=N)$ gauge groups, with $F_1$ and $F_2$ flavors. Seiberg duality on the $USp$ groups brings their ranks to $USp(N-M+F_1-4)$. A further duality on the $SU$ group brings it to $SU(N-2M+2F_1+F_2-8)$. Hence the quiver is self-similar with new values for $N$ and $M$ as
\begin{equation}
\label{cascf12}
N'=N-2M+2F_1+F_2-8\ , \qquad M'=M+4-F_1-F_2\ ,
\end{equation}
as shown in figure \ref{FQcas}.

Eq.~\eqref{cascf12} is in agreement with previous results. For $F_1=F_2=0$ we recover the case discussed in the previous section, the unflavored cascade, where $M$ increases towards the IR. Choosing instead $F_1+F_2=4$ the cascade enjoys constant steps and the two $\beta$-functions are just opposite one another, that is they sum to 0, as for the unflavored conifold and $\mathbf{F}_0$ cascades. Consistently, for $M=0$ we recover the SCFT discussed in section \ref{flconf}. For $F_1+F_2>4$, instead, things change. In particular, the number of fractional branes now {\it reduces} as we flow to the IR. The latter is the standard behavior for non-orientifolded duality cascades with flavors, as discussed for instance in \cite{Benini:2007gx}.

Starting at some UV scale $t_0$ from a configuration with $USp(N_0+M_0)$ and $SU(N_0)$ groups, after $k$ cascade steps one gets the ranks
\begin{equation}
N_k=N_0-2k M_0 -k(k+1)(4-F_1-F_2)-kF_2\ , \qquad M_k=M_0 +4-F_1-F_2\ .
\end{equation}
The $\beta$-functions at integral steps are 
\begin{align}
\beta^{USp}_k= 3M_k+\frac32(4- F_1) \quad , \quad \beta^{SU}_k= -3M_k-\frac32 F_2 \ ,
\end{align}
while at half steps
\begin{align}
{\beta^{USp}_k}= -3M_k-\frac32(4- F_1) \quad , \quad {\beta^{SU}_k}= 3M_k-\frac32 (2F_1+F_2-8)\ . 
\end{align}
Note that the sum of $\beta$-functions is always constant
\begin{equation}
\label{sumbetafl}
\beta^{USp}_k+ \beta^{SU}_k= \frac32(4- F_1-F_2)\ .
\end{equation}

Let us first consider the case $F_1+F_2=4$. In this case the value of $M$ is constant along the flow, $M_k=M_0\equiv M$, and the sum of the $\beta$-functions vanishes. The inverse couplings thus evolve, alternating at strong coupling, but always below a constant value. 

Specializing to the case $F_1=4$, the cascade is not just qualitatively but even quantitatively the same as the conifold and $\mathbf{F}_0$ cascade,s as far as the ranks of the gauge groups are concerned, that is  $N'=N-2M$ and $M'=M$. 

It is interesting to see what happens in the IR. If we assume that after $l$ steps, the $USp$ groups confine at the origin of the mesonic moduli space, we need to have $N_f=N_c+4$, which translates to $N_l=M$, that is $N_0=(2l+1)M$. We are then left with a pure $SU(M)$ SYM, which in turn confines, as in the parent conifold and $\mathbf{F}_0$ theories. This is a behavior usually associated to deformation branes. 

We can assume on the other hand that after $l$ (and a half) steps, it is the $SU$ group which finds itself on the baryonic branch. This happens when $N_l=2M$ and thus $N_0=(2l+2)M$. We are now left with two $USp(M)$ gauge groups with 4 flavors each. Supposing $M\geq 4$, a runaway superpotential is generated \cite{Intriligator:1995ne}. However, the runaway behavior is stopped by the surviving quartic superpotential. Indeed, we have
\begin{equation}
W_\mathrm{eff}= \mathcal{P}\tilde{\mathcal{P}} + \left(\frac{\Lambda_1^{\frac32M-2}}{\mathrm{Pf}\mathcal{P}}\right)^{\frac2{M-4}}+ \left(\frac{\Lambda_3^{\frac32M-2}}{\mathrm{Pf}\tilde{\mathcal{P}}}\right)^{\frac2{M-4}}\ ,
\end{equation}
where the mesons ${\mathcal P}$ and $\tilde{\mathcal{P}}$  are defined in eqs.~\eqref{mes1} and \eqref{mes2}. The mesons are massive, and it is straightforward to see that there are solutions to the F-term equations, all with $\mathrm{Pf}\mathcal{P},\mathrm{Pf}\tilde{\mathcal{P}}\neq0$.

Conversely, one can consider the case with $F_2=4$. It is easy to see that, with respect to the previous case, the role of the symplectic and unitarity groups gets interchanged, as far as the IR end of the cascade is concerned. For later convenience, let us label the starting point as $USp(N_0+M-2)$ and $SU(N_0)$.  After $l$ steps we have $N_l=N_0-2lM$. If the $USp$ groups confine first, we have to set $N_l=M+2$ and so $N_0=(2l+1)M+2$. We are thus left with a $SU(M+2)$ with 4 flavors. As soon as $M>2$, an ADS superpotential \cite{Affleck:1983mk} is dynamically generated, but the runaway is again stopped by the quartic tree-level term
\begin{equation}
W_\mathrm{eff}= \mathcal{R}^2 + \left(\frac{\Lambda_2^{3M+2}}{\mathrm{det}\mathcal{R}}\right)^{\frac1{M-2}}\ .
\end{equation}
If after $l$ (and a half) steps we confine $SU$ on the baryonic branch, we need $N_l=2M$ and $N_0=(2l+2)M$. We end up with two decoupled pure $USp(M-2)$ SYM theories, which confine into supersymmetric vacua. 

It is now quite obvious that in the configuration $F_1=F_2=2$, the only possible situation at the bottom of the cascade is confinement with a stabilized ADS superpotential.

We can also contemplate the UV behavior of these cascades. It actually goes on indefinitely, as in the conifold and $\mathbf{F}_0$ parents, with  a constant value for the sum of the two inverse squared gauge couplings, and a regular increase in the overall ranks. The UV completion in this case is with an infinite number of degrees of freedom. This case is therefore the closest we get to the canonical conifold cascade, in other words the effect of the orientifold is fully compensated by the 4 flavor D7-branes.

Let us now consider the case with $F_1+F_2>4$ flavors. The sum of the $\beta$-functions \eqref{sumbetafl} is negative. Thus the inverse couplings run alternatingly under an envelope which is tilted downwards in the UV. At the same time, the effective number of fractional branes $M_k$ decreases in the IR while it increases in the UV. This is a situation very similar to the flavored conifold \cite{Benini:2007gx}. In the IR, the theory either confines in one of the usual ways, if there is still a sizeable value of $M$, or it goes to a phase in which at least one of the groups is IR free, while the other(s) could be at a conformal fixed point, if the effective $M$ goes to zero before. In the UV, one runs into a duality wall. Indeed, the envelope for the sum of the inverse couplings crosses the horizontal axis of infinite coupling at some UV scale, before which there is an infinite sequence of Seiberg dualities which make the overall ranks infinite in a finite RG time. This  pathological behavior has already been described in \cite{Benini:2007gx}.

\section{A different orientifold quiver}
\label{secnonchi}

In this section we take one step towards the generalization of the previous discussion, by taking the orientifold of a different quiver. We will start with a different $\mathbf{Z}_2$ orbifold of the conifold, namely a non-chiral one, which was extensively studied in \cite{Argurio:2006ny,Argurio:2007qk,Aharony:2007pr,Argurio:2008mt}. For our purposes, it will suffice to focus on the cascade without flavors, highlighting its main features, in view of a comparison with the previous chiral model.

The non-chiral  $\mathbf{Z}_2$ orbifold of the conifold is part of a family of quiver gauge theories, corresponding to $\mathbf{Z}_n$ orbifolds of the conifold, which are composed of a necklace of $2n$ $SU(N_i)$ gauge groups, and non-chiral sets of bifundamentals between each adjacent nodes. For $n=2$, which is what we are going to focus on, the theory has just four gauge factors. The matter fields being non-chiral means that there are no gauge anomalies that restrict the ranks of the gauge groups. Accordingly, there are three different kinds of fractional D3-branes, modulo regular ones. The interplay of such different kinds of fractional branes was studied in \cite{Argurio:2008mt}. 

\begin{figure}[ht]
\begin{center}
\includegraphics[height=0.25\textheight]{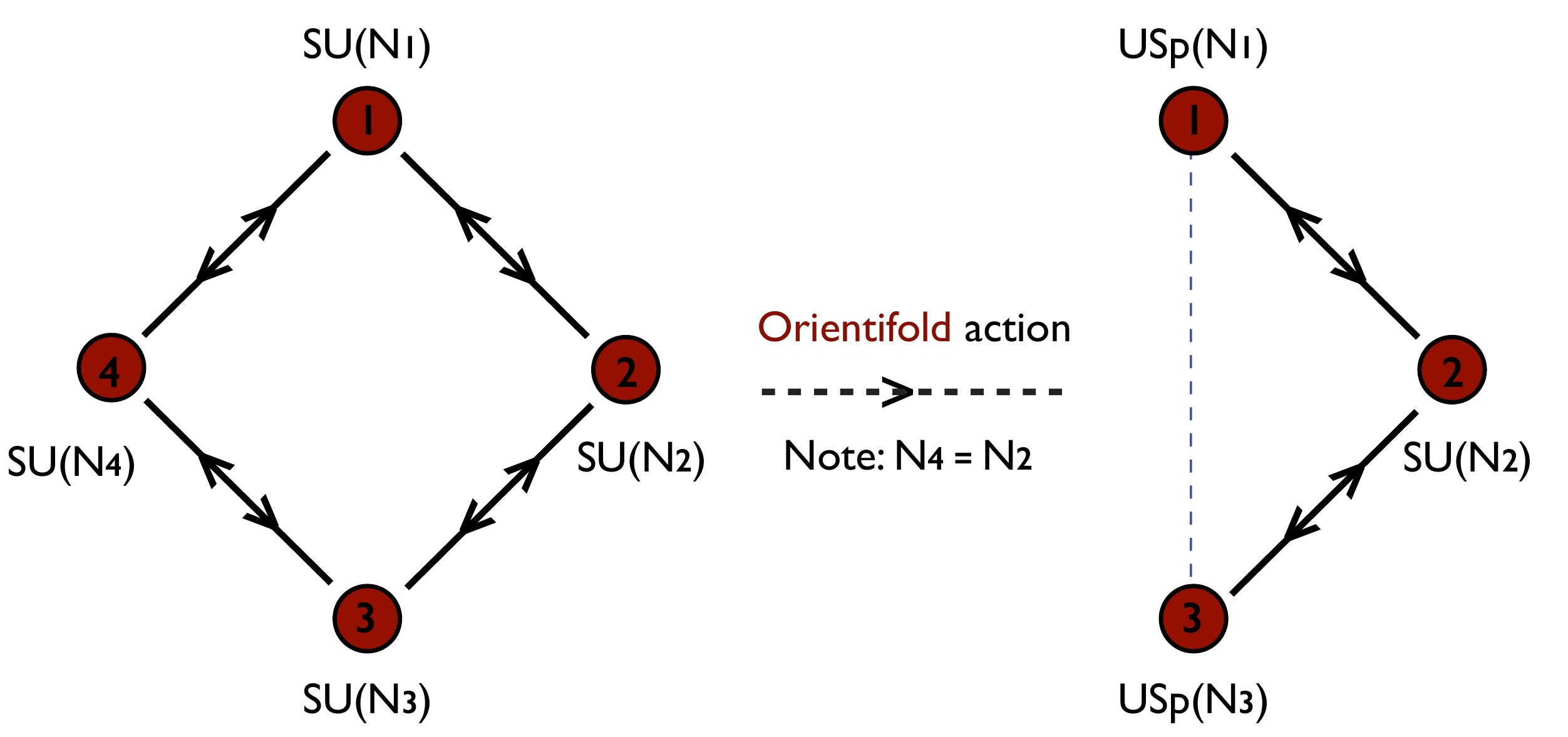}
\caption{The non-chiral $\mathbf{Z}_2$ orbifold of the conifold and its orientifold.}
\label{Z2nc}
\end{center}
\end{figure}

Upon orientifolding the necklace opens up, with the two end-nodes being  $USp$ while the $2n-2$ middle ones are identified pairwise and remain $SU$. Hence, in our $\mathbf{Z}_2$ case we have three nodes, with the first and third being symplectic and the second one unitary. The quiver, which is reported in figure \ref{Z2nc} together with the parent orbifold theory,  looks the same as the previous $\mathbf{F}_0/\Omega$ quiver, with the notable difference that the double bi-fundamental lines have now opposite arrows instead of pointing in the same direction. As a consequence, differently from $\mathbf{F}_0/\Omega$, the ranks of the two $USp$ nodes are free to be different. In other words, two fractional branes survive the orientifold projection. The superpotential is quartic and reads
\begin{equation}
W= X_{12}X_{23}X_{32}X_{21} - {X_{12}}^TX_{12}X_{21}{X_{21}} ^T- {X_{32}}^TX_{32}X_{23}{X_{23}}^T ~.
\end{equation}
Such quiver and superpotential have been shown to be self-similar under a cascade of Seiberg dualities (acting first on the $USp$ nodes, then on the $SU$ one) in \cite{Aharony:2007pr}.
Choosing the three ranks to be the same, $N_i=N$, we get a very similar picture as for the chiral case. In particular, using the same conventions as in that case, we have
\be
\label{noMfrnc}
\beta^{USp} = 6 > 0\ , \qquad \qquad \beta^{SU}= 0\ ,
\ee
which are the same as eqs.~\eqref{noMfr} of the chiral orientifold, and so is the UV dynamics. In what follows, we want to discuss the RG-flows which are generated adding one, respectively two kinds of fractional branes, on top of $N$ regular branes.

\subsection{Cascade with one type of fractional branes}

Let us first consider the case where the $USp$ nodes have equal ranks, $N_1$, but different from that of the $SU$ node, $N_2$. The $\beta$-functions can be easily computed to be 
\begin{align}
\beta^{USp}&= 3(N_1+2)-3N_2 \ , \\
\beta^{SU}&= 3N_2-3N_1 \ .
\end{align}
We see that, as for $\mathbf{F}_0/\Omega$, the sum of $\beta$-functions is always positive, $\beta^{USp}+\beta^{SU}=6$, so there is no conformal phase and, in the average, the couplings become stronger as the energy decreases. 

Let us take, for definiteness, $N_1=N+M$ and $N_2=N$. Dualities on the $USp$ nodes take $USp(N+M)$ to $USp(N-M-4)$, and a further Seiberg duality on the $SU$ node takes $SU(N)$ to $SU(N-2M-8)$. We thus see that after a cycle we have the new ranks
\begin{equation}
N'=N-2M-8\ , \qquad M'=M+4\ ,
\end{equation}
exactly as for $\mathbf{F}_0/\Omega$. We can thus transpose to the present case all the discussion about the RG flow, and the scale at which the groups confine. Indeed one can show, following for instance \cite{Aharony:2007pr}, that the steps through which one gets to either pure $SU(M)$ SYM, or a pair of decoupled $USp(M)$s, are also very similar to our previous example. We thus conclude that, also in this geometry, the effect of adding an orientifold is to allow for an RG flow which starts in the UV with a mix of conformal and/or mildly asymptotically free gauge groups, and ends in the IR with confinement after a cascade of Seiberg dualities which gets denser and denser towards a would-be IR duality wall. To summarize, this one-kind fractional branes case does not differ in any sensible manner from the chiral orientifold considered in section \ref{chunf} and shares with it its (interesting) RG-flows, like finite duality cascades leading to confinement.

\subsection{Cascade with two types of fractional branes}

We now  investigate the effect of having more than one type of fractional branes in business. Let us then consider a $USp(N+M) \times SU(N) \times USp(N+P)$ quiver gauge theory. The RG-flow is now governed by
\begin{equation}
\beta^{USp}_1= \frac32(M+2) \ , \qquad
\beta^{SU}_2= -\frac32(M+P) \ , \qquad
\beta^{USp}_3= \frac32(P+2) \ ,
\end{equation}
and after a cycle of Seiberg dualities, the ranks are mapped to $N'=N-M-P-8$, $M'=P+4$ and $P'=M+4$. 

\begin{figure}[ht]
\begin{center}
\includegraphics[height=0.45\textheight]{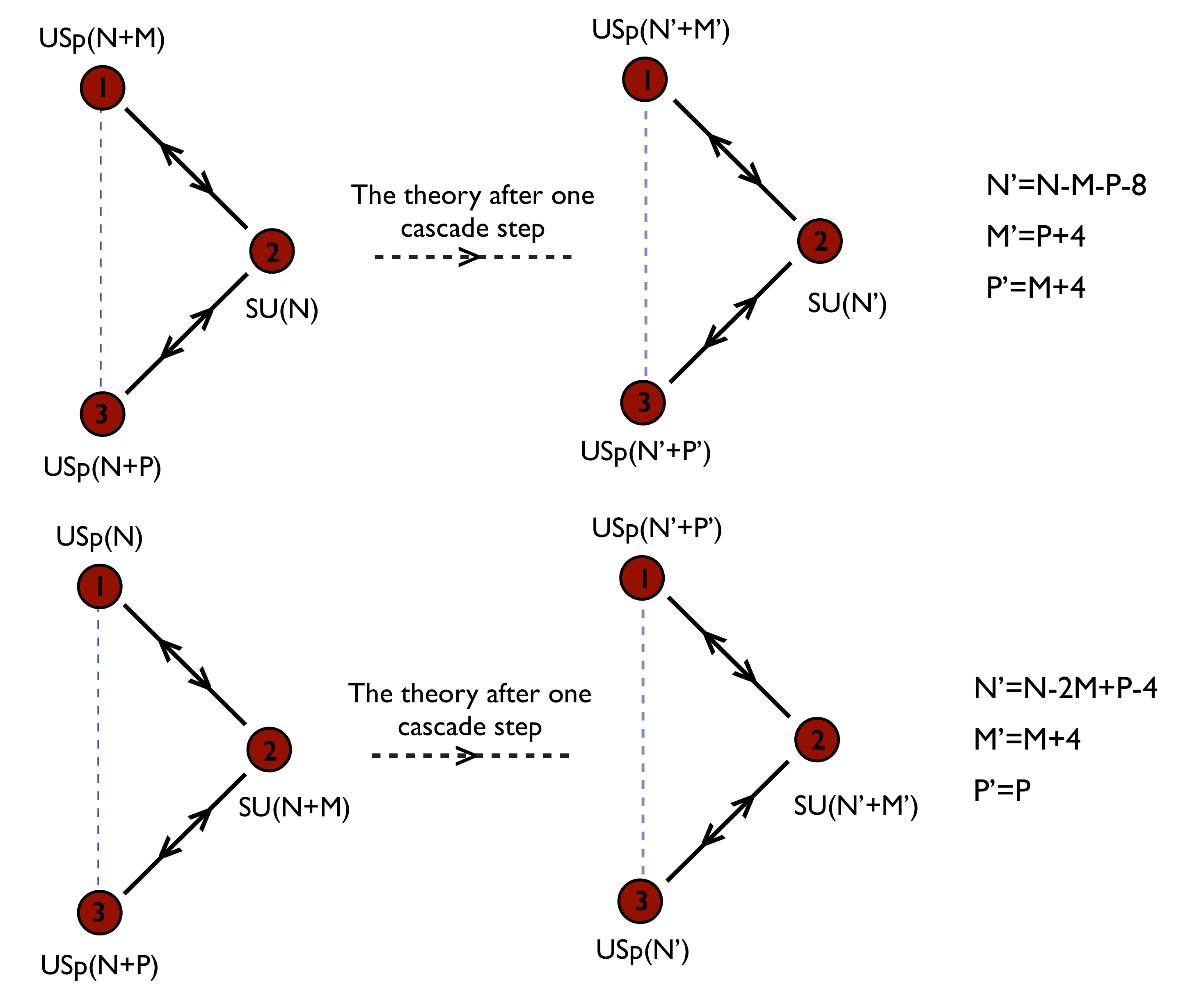}
\caption{The non-chiral orientifold quiver theories with two types of fractional branes and different rank assignments. Quivers on the right represent the theories after the  first cascade steps.}
\label{Z2ncfr}
\end{center}
\end{figure}

There is also a different possibility, which is to have the ranks ordered in a different way, with one of the $USp$ nodes having the smallest rank. For $USp(N)_1\times SU(N+M)_2\times USp(N+P)_3$, the $\beta$-functions are
\begin{equation}
\beta^{USp}_1=- \frac32(M-2) \ , \qquad
\beta^{SU}_2= \frac32(2M-P) \ , \qquad
\beta^{USp}_3= \frac32(P-M+2) \ .
\end{equation}
After a cycle of Seiberg dualities, the ranks are mapped to $N'=N-2M+P-4$, $M'=M+4$ and $P'=P$ (upon exchanging the roles of the two $USp$ nodes). The two cases are reported in figure \ref{Z2ncfr}.

The behavior at the bottom of the cascade depends on the relative sizes of the ranks. In the parent theory one can have either confinement, an effective $\mathcal{N}=2$ Coulomb branch or dynamical supersymmetry breaking with runaway behavior, as detailed in \cite{Argurio:2008mt}. Here, we would like to see which of these different behaviors survive the orientifold projection. In particular, we will try to recover the runaway behavior.\footnote{We do not expect to recover $\mathcal{N}=2$ Coulomb branch vacua. Indeed, in the parent four-node quiver, they are associated to having at the bottom of the cascade only two contiguous nodes with the same type of gauge groups (and same ranks). After the orientifold projection, this is no longer possible, since contiguous nodes have necessarily gauge groups of different types, $USp$ or $SU$.}

Typically, out of the three gauge groups, one will confine first, in a situation like $N_f=N_c$ for $SU$ or $N_f=N_c+4$ for $USp$. If $SU$ confines first, we are left with two decoupled $USp$ nodes, generically of different ranks, but which will confine exactly as in the equal ranks case. It is more interesting if one of the $USp$ nodes confines first. In this case we are left with a two-node quiver, with gauge groups $USp(M)\times SU(P)$ and superpotential 
\begin{equation}
W=-{X_{12}}^TX_{12}X_{21}{X_{21}}^T+\det X_{21}X_{12}\ .
\end{equation}
The second term is what is left over from the other $USp$ node, after it confined at the origin of moduli space. Alternatively, it can be reconducted to a stringy instanton on that node \cite{Argurio:2007qk,Aharony:2007pr}. Note that this term involves the determinant of a $P\times P$ matrix. If $M<P$, the matrix $X_{21}X_{12}$ is not of maximal rank, and the determinant term hence vanishes. In such a case, the $SU$ node is in a situation where $N_f<N_c$, and an Affleck-Dine-Seiberg superpotential is generated. In terms of the mesons $\mathcal{M}=X_{12}X_{21}$ of the $SU$ group the effective superpotential becomes
\begin{equation}
W_\mathrm{eff}=-\mathcal{M}\mathcal{M}^T+\left(\frac{\Lambda_2^{3P-M}}{\det \mathcal{M}}\right)^\frac{1}{P-M}\ .
\end{equation}
Such superpotential has SUSY vacua.

When $M\geq 2P$, on the other hand, the $USp$ node is such that $N_f\leq N_c$ and it generates its own ADS superpotential.\footnote{The case $P\leq M<2P$ can be mapped to the present case after a further Seiberg duality on the $SU$ node.} In terms of the $USp$ mesons $\mathcal{N}=X_{21}X_{12}$, $\mathcal{A}=X_{12}^TX_{12}$ and 
$\tilde{\mathcal{A}}=X_{21}X_{21}^T$, which assemble into 
\begin{equation}
\mathcal{P}=
\left(
\begin{array}{cc}
\mathcal{A}  & \mathcal{N}    \\
-\mathcal{N}  &       \tilde{ \mathcal{A}}
\end{array}
\right)\ ,
\end{equation}
the effective superpotential reads
\begin{equation}
W_\mathrm{eff}=-\mathcal{A}\tilde{\mathcal{A}}+\det \mathcal{N}+\left(\frac{\Lambda_1^{\frac12(3M-2P+6)}}{\mathrm{Pf}\mathcal{P}}\right)^\frac{2}{M-2P+2}\ .
\end{equation}
One can see that all potentially runaway directions are again blocked, either by the mass term $\mathcal{A}\tilde{\mathcal{A}}$ or by the $\det \mathcal{N}$ term. Note that the latter term is specific to the orientifold quiver. We thus conclude that the runaway directions of the parent theory are stabilized in a supersymmetric way by the orientifold projection. Of all diverse behaviors of the parent theory, only the confining behavior remains an option. This is in nice agreement with the findings in \cite{Argurio:2008mt}. There it was shown that the runaway corresponds to the fractional branes associated to the ${\cal N}=2$ Coulomb branch (the so-called ${\cal N}=2$ fractional branes) being pushed all the way to infinity but, as already observed, such degrees of freedom are projected out by the orientifold.

\section{Discussion}
\label{disc}

As compared to their parent orbifold theories, the orientifolds models we have discussed have the following two, very peculiar properties. 

The first is that duality cascades are finite. This has the beautiful outcome that, unlike ordinary quiver gauge theories, cascading orientifold theories do not require an infinite number of degrees of freedom to be UV completed. Moreover, properly choosing the number of regular and fractional branes at the singularity, these finite RG-flows can interpolate between confining vacua and asymptotically free UV fixed points. 

The second interesting property is that orientifolds typically do not lead to runaway behavior. And, if runaway directions are present in the parent theory, orientifolds tend to stabilize them, as we have seen in the second model we discussed.

We have  analyzed two models in detail, one chiral and one non-chiral, but  the above properties are more tied to the presence of O-planes themselves rather than to the specific parent theory one starts with. Hence, we believe these to be generic properties of orientifold quiver theories.\footnote{Of course, it is possible to imagine a more complicated singularity, which gives rise to a quiver with many nodes, where only a tiny fraction of it is affected by the O-plane. For instance, in orientifolds of non-chiral $\mathbf{Z}_n$ orbifolds of the conifold discussed in section \ref{secnonchi}, for $n$ large, the middle of the chain will contain nodes associated to fractional branes which are not affected by the presence of the O-plane. We are interested in the physics of the nodes nearby the O-plane.}  

There are two aspects related to our discussion to which the properties summarized above might be of some relevance, and which we would like to comment upon. The first is  the possibility to have a weakly coupled gravity dual for orientifold cascades from the IR all the way up to the UV. The second aspect regards the possibility, discussed in, {\it e.g.}, \cite{Franco:2007ii,Retolaza:2015nvh}, that orientifold quiver gauge theories might admit stable (in addition to metastable \cite{Argurio:2007qk}) dynamically generated supersymmetry breaking vacua.

\subsection{Comments on the gravity dual}

A first obvious question is whether a gravity dual description of orientifold RG flows exists. Naively, orientifold effects appear to be $1/N$ suppressed, and it is known that such effects are usually considered to be captured holographically only outside the classical supergravity regime. In the present case, there is however one sharp difference: subleading corrections pile up in such a way that they become leading at the IR end of the cascade, hence making it possible for orientifold effects to have a visible backreaction also at the supergravity level. Let us see how this might come about.

As we have seen, an orientifold cascade is specified by the rank in the deep UV, and $1/g^2$, the sum of the inverse gauge couplings squared,  which, along the cascade, has a subleading in $1/N$ evolution. For a gravity dual to exist, the minimal requirement is that both $N\gg1$ and $g^2N\gg1$. In the deep UV, both $M$, the number of fractional branes, and the $\beta$-functions are zero to leading order in $N$, so that the theory is approximately conformal. Thus we expect to have a gravity dual which, to a good approximation and for large enough radius, has an $AdS_5$ factor times a compact 5-manifold, with $N$ units of $F_5$ flux. Of course, in the deep UV there might be a perturbation due to the $1/N$ running of some couplings to asymptotic freedom. However, this would happen after a parametrically large radial running such that the RG-time $t$ is proportional to $N$ and $g^2N$ becomes of order one. Going to the IR, the effective number of regular branes, $N$, decreases while $M$ increases. This means that going towards the interior of the geometry along the radial direction, the $F_5$ flux decreases while the $F_3$ flux starts increasing. Eventually, the geometry should cap off, with a vanishing $F_5$ flux and a finite $F_3$ flux. Let us focus, for definiteness, on the chiral orientifold cascade discussed in section \ref{chunf}. The $F_3$ flux is given by $M\simeq 2\sqrt{N}$. If $M\gg1$ and $g^2M\gg1$, the gravity dual should still be reliable.\footnote{Note that the couplings at the bottom and the top of a cascade consisting in $l$ steps are related by $g_b^2=l g_t^2$, cfr.~\eqref{rune0}. Hence the 't Hooft coupling at the bottom of the cascade reads $g_b^2M =g_t^2 lM\simeq g_t^2 M^2/4 \simeq g_t^2 N$, the same as the one at the top of the cascade.} This suggests that the gravity dual should be able to grasp the piling up of $1/N$ effects from the orientifold to finally produce a backreaction, interpreted as a large $M$ effect. Therefore, quite interestingly, one could describe the full cascade, even though the UV-completion is in terms of a finite number of degrees of freedom. This is certainly something worth exploring more closely. Work is in progress in this direction \cite{RM2}.

A different issue related to $1/N$ corrections has to do with the (im)possibility to decouple the IR physics from the UV completion. For instance, in the prototypical conifold cascade \cite{Klebanov:2000hb}  it is only by taking into account $1/N$ corrections that one can zoom into the last cascade step and decouple the infinitely many degrees of freedom becoming relevant when running the cascade upwards.\footnote{
Analogously, in models where one engineers gauge theories by wrapping higher dimensional branes on topologically non-trivial cycles of smooth CYs, as in \cite{Maldacena:2000yy}, KK modes decouple at subleading order in $1/N$.} Our model will have the same problem, in that the cascade steps  will be compressed in the supergravity limit, so that it will be impossible to isolate a pure SYM regime.

\subsection{Comments on stable supersymmetry breaking vacua}

Another aspect which makes orientifold field theories engaging has to do with supersymmetry breaking. It is known that in quiver gauge theories without orientifolds, one can, at best, find cascades that end up with supersymmetry breaking with runaway behavior \cite{Berenstein:2005xa,Franco:2005zu,Bertolini:2005di,Intriligator:2005aw} or, possibly, metastable vacua \cite{Kachru:2002gs,Argurio:2006ny}. The introduction of orientifolds has already been proven to be the most promising scenario to provide concrete realizations of metastable supersymmetry breaking vacua in string theory \cite{Argurio:2007qk}. Even more strikingly, orientifolds have been argued to provide quiver gauge theories which, at the bottom of the cascade, could lead to {\it stable} dynamically generated supersymmetry breaking vacua \cite{Franco:2007ii,Retolaza:2015nvh}.\footnote{A selection of further references where (meta)stable supersymmetry breaking is achieved with the help of orientifolds is \cite{Aharony:2007db,Garcia-Etxebarria:2015lif,Retolaza:2016alb}.} In the latter case, one engineers a quiver such that, when only few fractional branes are left, a known model of stable dynamical supersymmetry breaking is reproduced. These usually imply the presence of either $USp$ or $SO$ groups, or $SU$ groups with matter in the (anti)symmetric representation. Both instances require orientifolds to be generated through D-brane configurations. Their interest lies in being an alternative to supersymmetry breaking through anti-branes.

In light of our results, it would be interesting to investigate the UV completion of such set ups, namely from the point of view of the cascade, as done in \cite{Retolaza:2015nvh}. As we have seen, subtle effects such as left-over higher dimension terms in the superpotential, or the change in the number of fractional branes along the flow, can affect the physics at low energies.

It could also be interesting to see what happens on mesonic branches higher up in the cascade. When a cascade ends with confinement, the physics of the mesonic branches is similarly the one of confinement at each point of the larger moduli space representing the motion of regular branes away from the singularity \cite{Dymarsky:2005xt}. When the endpoint is runaway, instead, it was shown that also mesonic branches become runaway in their own respect \cite{Argurio:2007vq}. Hence, one could wonder what happens to mesonic branches if at the bottom of the cascade there exist stable supersymmetry breaking vacua: supersymmetric moduli spaces, runaway, or other stable supersymmetry breaking vacua? Any one of these options would be much relevant to assess the global stability of the vacuum at the end of the cascade.

\section*{Acknowledgements}

We would like to thank Francesco Benini and Sergio Benvenuti for collaboration at the early stage of this project and for useful discussions all along. We are also grateful to Francesco Bigazzi, Cyril Closset, Aldo Cotrone, Sebastian Franco, Inaki Garcia-Etxebarria and Flavio Porri for discussions and feedback. Finally, we warmly thank the Florence Theory group for exceptional hospitality during the completion of this work. This research has been supported in part by IISN-Belgium (convention 4.4503.15) and by the MIUR-PRIN Project ``Non-perturbative Aspects Of Gauge Theories And Strings'' (Contract 2015MP2CX4). R.A. is a Research Director of the Fonds de la Recherche Scientifique-F.N.R.S. (Belgium). 


\end{document}